\documentclass[reprint, longbibliography, noeprint, superscriptaddress]{revtex4-1}

\usepackage[utf8]{inputenc}

\usepackage[lining,semibold]{libertine} 
\usepackage[libertine, cmintegrals, bigdelims, vvarbb]{newtxmath}

\usepackage{amsmath}
\usepackage{amsfonts}
\usepackage{mathrsfs}
\usepackage{gensymb}
\usepackage{bbm}
\usepackage{dsfont}

\usepackage{kbordermatrix}
\renewcommand{\kbldelim}{(}
\renewcommand{\kbrdelim}{)}

\usepackage{chemformula}
\usepackage[caption=false]{subfig}
\usepackage{chemfig}
\usepackage[version=3]{mhchem}

\usepackage{tikz}
\usetikzlibrary{matrix,positioning,decorations.pathreplacing}

\usepackage{soul}
\usepackage{xcolor}

\usepackage{scalerel} 
\usepackage{comment}

\usepackage{blkarray}

\definecolor{webgreen}{rgb}{0,.5,0}
\definecolor{webbrown}{rgb}{.6,0,0}
\definecolor{grigio}{rgb}{.85,.85,.85} 
\definecolor{RoyalBlue}{rgb}{0.0, 0.14, 0.4}
\definecolor{skyblue1}{rgb}{0.45,0.62,0.81}
\definecolor{skyblue2}{rgb}{0.2,0.39,0.64}
\definecolor{skyblue3}{rgb}{0.13,0.29,0.53}
\definecolor{scarlet1}{rgb}{0.93,0.16,0.16}
\definecolor{scarlet2}{rgb}{0.8,0,0}
\definecolor{scarlet3}{rgb}{0.64,0,0}

\definecolor{g}{gray}{0.50}

\usepackage{hyperref}
\hypersetup{%
    colorlinks=true, linktocpage=true, pdfstartpage=1, pdfstartview=FitV,%
    breaklinks=true, pdfpagemode=UseNone, pageanchor=true, pdfpagemode=UseOutlines,%
    plainpages=false, bookmarksnumbered, bookmarksopen=true, bookmarksopenlevel=1,%
    hypertexnames=true, pdfhighlight=/O,
    urlcolor=webbrown, linkcolor=RoyalBlue, citecolor=webgreen, 
    pdftitle={},%
    pdfauthor={Francesco Avanzini},%
    pdfsubject={},%
    pdfkeywords={},%
    pdfcreator={pdfLaTeX},%
    pdfproducer={LaTeX REVTeX}%
}

\usepackage{cleveref}

\newenvironment{proof}{\textit{\textbf{Proof.}}}{\hfill$\square$}

\newcommand{\rem}[1]{{\color{black}#1}}
\newcommand{\cmark}[1]{{\color{black}#1}}

\newcommand{\remark}{\textit{\rem{Remark.} }}
\newcommand{\assumption}{\textit{{Assumption.} }}

\newcommand{\effp}[1]{\hat{#1}}
\newcommand{\actual}[1]{{#1}}

\newcommand{\abs}[1]{|#1|}

\newcommand{\dt}{\mathrm{d}_t}
\newcommand{\pt}{\partial_t}
\newcommand{\dd}{\mathrm d}

\newcommand{\dk}{{\mathbb 1}}

\newcommand{\rank}{\mathrm{rank}}

\newcommand{\mini}{*}
\newcommand{\minii}{\bullet}

\newcommand{\pos}{\boldsymbol{r}}

\newcommand{\ds}{\boldsymbol{\nabla}}

\newcommand{\setspe}{\mathcal S}
\newcommand{\nspe}{n_{s}}

\newcommand{\setspex}{X}
\newcommand{\setspexi}{X_I}
\newcommand{\setspexd}{X_D}

\newcommand{\setspey}{Y}
\newcommand{\setspeyp}{Y_p}
\newcommand{\setspeyf}{Y_f}

\newcommand{\nr}{\text{nr}}

\newcommand{\setspeyi}{\tilde{Y}}

\newcommand{\nspex}{n_\spex}
\newcommand{\nspey}{n_\spey}

\newcommand{\setcom}{\mathcal K}
\newcommand{\ncom}{n_{k}}

\newcommand{\setrct}{\mathcal R}
\newcommand{\nrct}{n_{r}}

\newcommand{\setrcte}{\mathcal R_{\text{ex}}}
\newcommand{\nrcte}{n_\spey}

\newcommand{\spe}{i}
\newcommand{\spee}{j}

\newcommand{\spex}{x}
\newcommand{\spexx}{{x'}}
\newcommand{\spey}{y}

\newcommand{\speyi}{\tilde{\spey}}

\newcommand{\speyp}{{y_p{}}}
\newcommand{\speypp}{{y_p{}'}}
\newcommand{\speyf}{{y_f{}}}

\newcommand{\rct}{\rho}
\newcommand{\rcty}{\rho_{\spey}}

\newcommand{\com}{k}

\newcommand{\chem}{Z_{\spe}}
\newcommand{\chemx}{Z_{\spex}}
\newcommand{\chemy}{Z_{\spey}}
\newcommand{\chemyi}{Z_{\speyi}}

\newcommand{\comp}[1]{V_{\com#1}}

\newcommand{\mi}{m_{\spe}}

\newcommand{\stoc}[2]{\nu_{#2, #1\rct}}

\newcommand{\stocx}[1]{{\nu}_{\spex, #1\rct}}
\newcommand{\stocy}[1]{{\nu}_{\spey, #1\rct}}

\newcommand{\stocxc}[1]{{\nu}_{\spex, \com#1}}

\newcommand{\matS}{\mathbb{S}}

\newcommand{\matSX}{\mathbb{S}_{\setspex}}
\newcommand{\matSY}{\mathbb{S}_{\setspey}}

\newcommand{\matSi}[1]{{S}_{\spe,#1\rct}}
\newcommand{\matSx}[1]{{S}_{\spex,#1\rct}}
\newcommand{\matSy}[1]{{S}_{\spey,#1\rct}}
\newcommand{\matSyy}[1]{{S}_{\spey,#1\rct'}}
\newcommand{\matSs}[2]{{S}_{#2,#1\rct}}

\newcommand{\matI}{\partial}
\newcommand{\matIe}{\partial_{\com,\rct}}
\newcommand{\matC}{\Gamma}
\newcommand{\matCe}[1]{\Gamma_{\spex,\com#1}}

\newcommand{\deff}{\delta}

\newcommand{\setpth}{\mathcal{E}}
\newcommand{\npth}{n_{e}}

\newcommand{\pth}{\varepsilon}
\newcommand{\setrctpth}[1]{\setrct_{#1\pth}}

\newcommand{\stocxe}[1]{{\nu}_{\spex, #1\pth}}

\newcommand{\matSSxe}{\effp{\mathbb{S}}_{\setspex}}
\newcommand{\matSSxie}{\effp{\mathbb{S}}_{\setspexi}}
\newcommand{\matSxe}[1]{{S}_{\spex,#1\pth}}

\newcommand{\cyclev}{\boldsymbol{\phi}}
\newcommand{\cycle}{\phi_{\rct}}
\newcommand{\cyclee}{\phi_{\pth}}

\newcommand{\conc}{c_\spe(\pos)}

\newcommand{\concm}{c_{\spe}^{\mini}(\pos)}
\newcommand{\conxm}{c_{\spex}^{\mini}(\pos)}
\newcommand{\conxmm}{c_{\spex}^{\minii}(\pos)}

\newcommand{\conv}{\boldsymbol{c}(\pos)}
\newcommand{\convv}{\boldsymbol{c}}

\newcommand{\conveq}{\boldsymbol{c}_{\mathrm{eq}}(\pos)}
\newcommand{\convveq}{\boldsymbol{c}_{\mathrm{eq}}}
\newcommand{\convss}{\boldsymbol{c}_{\mathrm{ss}}(\pos)}
\newcommand{\convvss}{\boldsymbol{c}_{\mathrm{ss}}}

\newcommand{\convm}{\boldsymbol{c}^{\mini}(\pos)}
\newcommand{\convmm}{\boldsymbol{c}^{\minii}(\pos)}

\newcommand{\conveqh}{\boldsymbol{c}_{\mathrm{eq}}^{\mathrm{h}}}
\newcommand{\convssh}{\boldsymbol{c}_{\mathrm{ss}}^{\mathrm{h}}}

\newcommand{\conx}{c_\spex^{}(\pos)}
\newcommand{\conxx}{c_\spexx^{}(\pos)}
\newcommand{\conxh}{c_\spex^{\mathrm{h}}}

\newcommand{\conxss}{c_\spex^{\mathrm{ss}}(\pos)}

\newcommand{\cony}{c_\spey^{}(\pos)}
\newcommand{\conyy}{c_\spey^{}}

\newcommand{\dcurr}{\boldsymbol{J}_{\spe}(\conv)}
\newcommand{\dcurreq}{\boldsymbol{J}_{\spe}(\conveq)}
\newcommand{\dcurrss}{\boldsymbol{J}_{\spe}(\convss)}

\newcommand{\dcurrx}{\boldsymbol{J}_{\spex}(\conv)}

\newcommand{\matO}{\mathbb{O}(\conv)}
\newcommand{\matOij}{O_{\spe,\spee}(\conv)}

\newcommand{\Di}{D_\spe}
\newcommand{\Dx}{D_\spex}

\newcommand{\rcurr}[1]{j_{#1\rct}(\conv)}
\newcommand{\rcurreq}[1]{j_{#1\rct}(\conveq)}
\newcommand{\rcurrss}[1]{j_{#1\rct}(\convss)}

\newcommand{\rcurrssh}[1]{j_{#1\rct}(\convssh)}

\newcommand{\rflux}[1]{\omega_{#1\rct}(\conv)}
\newcommand{\rfluxx}{\omega_{\mathrm{in}}(\conv, \{\stoc{}{\spex}\})}

\newcommand{\rfluxyy}{\omega_{\mathrm{ch}}(\{\conyy\}, \{\stoc{}{\spey}\})}
\newcommand{\kconst}[1]{k_{#1\rct}}

\newcommand{\sr}[1]{s_{#1\rct}(\conv)}

\newcommand{\ecurr}{I_{\spe}(\pos)}
\newcommand{\ecurrx}{I_{\spex}(\pos)}
\newcommand{\ecurry}{I_{\spey}(\pos)}
\newcommand{\ecurryp}{I_{\speyp}(\pos)}
\newcommand{\ecurrypp}{I_{\speypp}(\pos)}
\newcommand{\ecurryf}{I_{\speyf}(\pos)}

\newcommand{\pcurr}[1]{\effp{j}_{#1\pth}(\conv)}

\newcommand{\pflux}[1]{\effp{\omega}_{#1\pth}(\conv)}

\newcommand{\Ap}[1]{a_{#1\rct}}
\newcommand{\AAp}[1]{A_{#1\rct}}

\newcommand{\aap}[1]{a_{#1\pth}(\conv)}
\newcommand{\bbp}[1]{b_{#1\pth}}

\newcommand{\setilaw}{\Lambda}
\newcommand{\ilaw}{\lambda}

\newcommand{\setilawb}{\Lambda_b}
\newcommand{\ilawb}{\lambda_b}
\newcommand{\setilawu}{\Lambda_u}
\newcommand{\ilawu}{\lambda_u}

\newcommand{\law}[2]{{\ell}^{#1}_{#2}}

\newcommand{\claw}{{\ell}^{\ilaw}_{\spe}}

\newcommand{\clawv}{\boldsymbol{\ell}^{\ilaw}}

\newcommand{\clawmv}{\boldsymbol{\ell}^{m}}

\newcommand{\cqua}{L^{\ilaw}[\convv]}
\newcommand{\cquaref}{\actual{L}^{\ilaw}}

\newcommand{\clawb}{{\ell}^{\ilawb}_{\spe}}
\newcommand{\clawvb}{\boldsymbol{\ell}^{\ilawb}}
\newcommand{\cquab}{L^{\ilawb}[\convv]}
\newcommand{\clawbx}{{\ell}^{\ilawb}_{\spex}}
\newcommand{\clawby}{{\ell}^{\ilawb}_{\spey}}
\newcommand{\clawbp}{{\ell}^{\ilawb}_{\speyp}}
\newcommand{\clawbpp}{{\ell}^{\ilawb}_{\speypp}}
\newcommand{\clawbf}{{\ell}^{\ilawb}_{\speyf}}
\newcommand{\iclawbp}{\overline{\ell}{}_{\ilawb}^{\speyp}}
\newcommand{\iclawbpp}{\overline{\ell}{}_{\ilawb}^{\speypp}}
\newcommand{\cmoi}{M_{\speyp}[\convv]}
\newcommand{\clawu}{{\ell}^{\ilawu}_{\spe}}
\newcommand{\clawux}{{\ell}^{\ilawu}_{\spex}}

\newcommand{\cquau}{L^{\ilawu}[\convv]}
\newcommand{\cquauref}{\actual{L}^{\ilawu}}

\newcommand{\clawbpex}{{\ell}^{\ilawb}_{\ch{Y_1}}}

\newcommand{\Fd}[1]{\frac{\delta#1}{\delta\conc}}
\newcommand{\Fdx}[1]{\frac{\delta#1}{\delta\conx}}

\newcommand{\free}{F[\convv]}

\newcommand{\freeid}{F^{\mathrm{id}}[\convv]}

\newcommand{\freeni}{F^{\mathrm{ni}}[\convv]}

\newcommand{\sgfree}{\mathcal{F}[\convv]}
\newcommand{\sgfreeeq}{\mathcal{F}[\convveq]}

\newcommand{\cp}{\mu_\spe(\conv)}
\newcommand{\cpid}{\mu^{\mathrm{id}}_\spe(\conc)}
\newcommand{\stcp}{\mu_\spe^\circ}
\newcommand{\act}{\gamma_\spe(\conv)}

\newcommand{\cpex}{\mu_\spe^{\mathrm{ex}}(\conv)}

\newcommand{\cpg}[1]{\mu_{#1}(\conv)}

\newcommand{\cpgss}[1]{\mu_{#1}(\convss)}
\newcommand{\cpgssh}[1]{\mu_{#1}(\convssh)}

\newcommand{\cpeq}{\mu_\spe(\conveq)}
\newcommand{\cpss}{\mu_\spe(\convss)}
\newcommand{\cpm}{\mu_\spe(\convm)}

\newcommand{\cpx}{\mu_\spex(\conv)}
\newcommand{\cpxm}{\mu_\spex(\convm)}
\newcommand{\cpxmm}{\mu_\spex(\convmm)}
\newcommand{\stcpx}{\mu_\spex^\circ}
\newcommand{\stcpy}{\mu_\spey^\circ}

\newcommand{\cpy}{\mu_\spey(\conv)}
\newcommand{\cpyy}{\mu_\spey}

\newcommand{\cpry}{\mu_\spey(\pos)}

\newcommand{\cpypeq}{\mu_{\speyp}(\conveq)}
\newcommand{\cpyfeq}{\mu_{\speyf}(\conveq)}

\newcommand{\cpryp}{\mu_{\speyp}(\pos)}
\newcommand{\cprypref}{\mu_{\speyp}^{\mathrm{ref}}}

\newcommand{\cpryf}{\mu_{\speyf}(\pos)}

\newcommand{\epr}{\dot{\Sigma}[\convv]}
\newcommand{\eprss}{\dot{\Sigma}[\convvss]}
\newcommand{\deprss}{\dot{\sigma}^{\mathrm{ss}}(\pos)}

\newcommand{\eprdff}{\dot{\Sigma}_{\mathrm{dff}}[\convv]}
\newcommand{\deprdff}{\dot{\sigma}_{\mathrm{dff}}(\pos)}
\newcommand{\deprdffss}{\dot{\sigma}^{\mathrm{ss}}_{\mathrm{dff}}(\pos)}
\newcommand{\eprrct}{\dot{\Sigma}_{\mathrm{rct}}[\convv]}
\newcommand{\deprrct}{\dot{\sigma}_{\mathrm{rct}}(\pos)}
\newcommand{\deprrctr}{\dot{\sigma}_{\rct}(\pos)}
\newcommand{\deprrctss}{\dot{\sigma}^{\mathrm{ss}}_{\mathrm{rct}}(\pos)}
\newcommand{\deprrctrss}{\dot{\sigma}^{\mathrm{ss}}_{\rct}(\pos)}
\newcommand{\eprrcte}{\effp{\dot{\Pi}}_{\mathrm{rct}}[\convv]}

\newcommand{\eprdffeq}{\dot{\Sigma}_{\mathrm{dff}}[\convveq]}
\newcommand{\eprrcteq}{\dot{\Sigma}_{\mathrm{rct}}[\convveq]}

\newcommand{\eprdffss}{\dot{\Sigma}_{\mathrm{dff}}[\convvss]}
\newcommand{\eprrctss}{\dot{\Sigma}_{\mathrm{rct}}[\convvss]}
\newcommand{\eprrctess}{\effp{\dot{\Pi}}_{\mathrm{rct}}[\convvss]}

\newcommand{\ncforce}[1]{f^{\mathrm{nc}}_{#1}(\pos)}

\newcommand{\ncwrk}{\dot{W}_{\mathrm{nc}}[\convv]}
\newcommand{\dncwrk}{\dot{w}_{\mathrm{nc}}(\pos)}
\newcommand{\ncwrkss}{\dot{W}_{\mathrm{nc}}[\convvss]}
\newcommand{\dncwrkss}{\dot{w}^{\mathrm{ss}}_{\mathrm{nc}}(\pos)}

\newcommand{\drwrk}{\dot{W}_{\mathrm{driv}}[\convv]}

\newcommand{\matM}{\mathbb{M}}
\newcommand{\matMij}{M_{\spex,\spexx}}

\newcommand{\matK}{\mathbb{K}}
\newcommand{\matKij}{K_{\spex,\spexx}}

\newcommand{\lm}{f}
\newcommand{\lmv}{\boldsymbol{\lm}}
\newcommand{\lmvv}{\boldsymbol{\lm}}
\newcommand{\lmvm}{\boldsymbol{\lm}^{\mini}}
\newcommand{\lmvmm}{\boldsymbol{\lm}^{\minii}}

\newcommand{\lmcl}{\lm_{\ilaw}^{}}
\newcommand{\lmclm}{\lm_{\ilaw}^{\mini}}
\newcommand{\lmclu}{\lm_{\ilawu}^{}}

\newcommand{\lmclum}{\lm_{\ilawu}^{\mini}}

\newcommand{\lmx}{\lm_{\spex}}
\newcommand{\lmxmm}{\lm^{\minii}_{\spex}}

\newcommand{\lagc}{\mathcal L[\convv, \lmvv]}

\newcommand{\Fdlmcl}[1]{\frac{\delta#1}{\delta\lmcl}}

\newcommand{\Fdlmclu}[1]{\frac{\delta#1}{\delta\lmclu}}

\newcommand{\Fdlmx}[1]{\frac{\delta#1}{\delta{\lmx}}}

\newcommand{\shx}{\Delta_{\spex}}
\newcommand{\cpxe}{\effp{\mu}_\spex(\conv)}
\newcommand{\freepdb}{{F}_{\mathrm{pdb}}[\convv]}

\newcommand{\freecb}{{F}_{\mathrm{cb}}[\convv]}

\newcommand{\cpxssh}{\mu_\spex(\convssh)}

\newcommand{\eprrctc}{\dot{\Pi}_{\mathrm{cb}}[\convv]}
\newcommand{\eprrctcss}{\dot{\Pi}_{\mathrm{cb}}[\convvss]}

\newcommand{\freeshift}{{F}_{\mathrm{pdb/cb}}[\convv]}

\newcommand{\rna}{a}
\newcommand{\rnb}{b}

\newcommand{\clawve}{\boldsymbol{\ell}}
\newcommand{\cquae}{L[\convv]}
\newcommand{\cpe}[1]{\mu_{#1}(\pos)}
\newcommand{\cprefe}[1]{\mu^{\mathrm{ref}}_{#1}}
\newcommand{\ecurre}[1]{I_{#1}(\pos)}

\newcommand{\pathway}{lumped reaction}
\newcommand{\pathways}{lumped reactions}

\makeatletter
\def\maketag@@@#1{\hbox{\m@th\normalfont\normalsize#1}}
\makeatother

\DeclareMathAlphabet{\mathpzc}{OT1}{pzc}{m}{it}

\begin{document}

\title{Nonequilibrium Thermodynamics of Non-Ideal Reaction-Diffusion Systems:\\
Implications for Active Self-Organization}

\newcommand\unilu{\affiliation{Department of Physics and Materials Science, University of Luxembourg, L-1511 Luxembourg City, Luxembourg}}
\newcommand\unipdchem{\affiliation{Department of Chemical Sciences, University of Padova, Via F. Marzolo, 1, I-35131 Padova, Italy}}
\author{Francesco Avanzini}
\email{francesco.avanzini@unipd.it}
\unipdchem
\author{Timur Aslyamov}
\email{timur.aslyamov@uni.lu}
\unilu
\author{\'Etienne Fodor}
\email{etienne.fodor@uni.lu}
\unilu
\author{Massimiliano Esposito}
\email{massimiliano.esposito@uni.lu}
\unilu


\date{\today}

\begin{abstract}
We develop a framework describing the dynamics and thermodynamics of open non-ideal reaction-diffusion systems, which embodies Flory-Huggins theories of mixtures and chemical reaction network theories. 
Our theory elucidates the mechanisms underpinning the emergence of self-organized dissipative structures in these systems. 
It evaluates the dissipation needed to sustain and control them, discriminating the contributions from each reaction and diffusion process with spatial resolution. 
It also reveals the role of the reaction network in powering and shaping these structures.
We identify particular classes of networks in which diffusion processes always equilibrate within the structures, while dissipation occurs solely due to chemical reactions. The spatial configurations resulting from these processes can be derived by minimizing a kinetic potential, contrasting with the minimization of the thermodynamic free energy in passive systems. 
This framework opens the way to investigating the energetic cost of phenomena such as liquid-liquid phase separation, coacervation, and the formation of biomolecular condensates.
\end{abstract}

\maketitle


\section{Introduction\label{sec:intro}}

Self-organization of molecules in solution can produce complex spatio-temporal structures 
that range from self-assemblies to phase separations, encompassing stationary patterns and oscillations. 
It results from nonlinear effects which can be caused by a variety of mechanisms.

One possible mechanism is based on molecular interactions, typical of \textit{non-ideal} systems, which can lead to self-organization even at equilibrium.
When no external energy is provided to sustain self-organization, 
this mechanism is \textit{passive} and described by a dynamics exhibiting \textit{detailed balance}.
The theoretical description for this mechanism is typically provided by 
Cahn-Hilliard theory of spinodal decomposition~\cite{cahn1958free} and its variants~\cite{Bray1994}.
It can be further understood from a thermodynamic perspective through theories like that proposed by Flory and Huggins~\cite{Huggins1941,Flory1942}
as a relaxation towards equilibrium. 

Another mechanism involves multimolecular chemical reactions in \textit{open} systems, 
which can also lead to self-organization in the absence of interactions. 
Predicted by A. M. Turing~\cite{turing1952} and 
extensively studied by the Brussels School~\cite{Prigogine1971,Nicolis1977,Lefever2018},
its experimental validation took nearly four decades~\cite{castets1990experimental}.
From a thermodynamic viewpoint, 
this type of self-organization is maintained by the continuous consumption of energy~\cite{Falasco2018a, Avanzini2019a, avanzini2020a},
making the mechanism \textit{active} with dynamics that exhibits \textit{broken detailed balance}.
{While our focus is on molecules in solution, we note that this mechanism also applies to larger spatial structures, such as predator-prey distributions~\cite{segel1972dissipative, Wang:2019aa},
vegetation patterns~\cite{Rietkerk2008,Reynes2017},
and spiral patterns of galaxies~\cite{cross2009pattern}.
} 

Recent years have seen renewed interest in the study of self-organization due to its crucial role in biological systems, 
where biomolecular condensates~\cite{Banani:2017condensates, Lyon2021} 
resulting from liquid-liquid phase separation {form} membraneless compartments. 
These condensates have been observed to serve various functions,
including 
storage of molecules~\cite{Klosin2020aa},
enhancement of reaction rates by concentrating enzymes and substrates~\cite{Castellana2014aa, Peeples2020},
and maintenance of macromolecule folding states~\cite{Frottin2019aa}.

{Many studies have attempted to reproduce the dynamics and the concentration profiles 
of these condensates~\cite{demarchi2023enzyme, glotzer1995reaction,brauns2020phase, haas2021turing},
without characterizing the interplay between passive~\cite{berry2018physical} and active~\cite{weber2019physics, zwicker2022intertwined} mechanisms
in their formation and regulation.
Indeed, their nonequilibrium nature is not easily discernible from examining the self-organized spatial structures alone. 
Instead, it requires a more detailed analysis of both diffusion and reaction fluxes, and how these fluxes are related to energy consumption. 
This critical feature is known as \textit{thermodynamic consistency}.
}


Understanding the control of self-organized structures 
by {out-of-equilibrium} chemical reactions and the resulting dissipation of energy is crucial,
especially in the field of biology. 
{Achieving passive control is nearly impossible due to the difficulty in adjusting molecular interactions. 
Consequently, it is necessary to rely on active mechanisms.
However, since biosystems operate within a limited energy budget 
and energy consumption is crucial for almost all biological processes, efficient regulation is essential.}
Previous studies have 
explored the regulation of such structures through simple unimolecular reactions~\cite{Kirschbaum2021, bauermann2022energy} 
or examined their stability in multimolecular reactions~\cite{aslyamov2023nonideal},
yet a comprehensive framework to analyze the nonequilibrium aspects of self-organization is still lacking. 
This work addresses this gap by highlighting the critical role of the reaction network topology 
in defining the relationship between self-organized structures and their thermodynamics. 

We develop a general thermodynamic theory for open non-ideal reaction-diffusion systems 
by integrating the thermodynamic theories of both 
{non-ideal chemical reactions}~\cite{Othmer1976, Avanzini2021} 
and ideal reaction-diffusion systems~\cite{Falasco2018a, Avanzini2019a}.
This unified approach enables us to investigate the energetics of self-organized structures and the interplay between passive and active mechanisms.
We examine various classes of reaction networks.
For \textit{detailed balanced} ones (open or closed), 
we show that the dynamics always relaxes to equilibrium while minimizing a suitable thermodynamic potential which plays the role of a Lyapunov function that depends only on the energetics and not on the kinetics. In these passive systems, the resulting self-organized structures do not dissipate.
For \textit{pseudo-detailed balanced} reaction networks, we show that the dynamics can be mapped into a detailed-balanced dynamics minimizing a kinetic potential that we identify and that plays the role of a Lyapunov function.
Nevertheless, these systems are active and the resulting self-organized structures dissipate.
For \textit{complex balanced} reaction networks (characterized by Arrhenius-like reaction fluxes),
such a mapping does not exist, but the dynamics still minimizes a kinetic potential that we construct and that plays the role of a Lyapunov function. These systems are also active, and the self-organized structures they produce also dissipate.
Remarkably, even though these systems are active, diffusion processes equilibrate at steady state in both pseudo-detailed balanced and complex balanced networks. 
However, most active systems do not fall into these classes.
Nonetheless, we show that our theoretical framework can be used to systematically analyze the energetics and the dissipation produced by the various active processes generating the self-organized structures.

The paper is structured as follows.
We begin in Section~\ref{sec:setup} by outlining the basic setup. 
Specifically, we define the representations of chemical reactions in terms of chemical reaction networks and graphs of complexes,
which are crucial for the derivations of our results.
In Section~\ref{sec:dyn}, 
we introduce the dynamical description of the concentration fields
and explore the concept of detailed balance in Subsection~\ref{sub:ss}.
In Subsection~\ref{sub:claw}, 
we discuss the conservation laws of chemical reaction networks to identify molecule fragments, known as moieties, which are transferred between chemical species through reactions while remaining intact. 
These moieties will be instrumental in identifying the nonconservative forces that break detailed balance.
We turn to thermodynamics in Section~\ref{sec:thermo}. 
We first discuss the conditions needed for the dynamics to be thermodynamically consistent. 
We then use moieties to rewrite the second law of thermodynamics in a way that identifies the thermodynamic forces driving the system out of equilibrium and the thermodynamic potential acting as a Lyapunov function when forces are switched off.
In Section~\ref{sec:aps}, 
we make use of thermodynamics, along with the representations of chemical reactions in terms of chemical reaction networks and graphs of complexes, to derive our main findings for pseudo-detailed balanced networks and complex balanced networks.
We end our study in Section~\ref{sec:num} by analyzing the energetics of various reaction networks using numerical simulations, including a network that is neither pseudo-detailed balanced nor complex balanced.
Conclusions are drawn in Section~\ref{sec:conc}.


\section{Basic Setup\label{sec:setup}}

We consider non-ideal reaction-diffusion (RD) systems in solution:
$\nspe$ interacting, reacting, and diffusing chemical species (identified by the labels $\spe\in\setspe$) are mixed together 
with a non-reacting and abundant species called solvent
which maintains the temperature $T$ and the volume $V$ constant.
We further assume that RD systems are embedded within impermeable walls.
The species $\setspe$ are
either chemostatted (labeled $\spey\in\setspey$), 
if exchanged with external reservoirs called chemostats (labeled $\spey\in\setspey$ too),
or internal (labeled $\spex\in\setspex$).
We use $\nspey$ and $\nspex$ for the number of chemostatted and internal species, respectively.
Due to the exchanges with the chemostats, RD systems are said to be open.
Without exchanges, RD systems would be closed.

\subsection{Chemical Reaction Networks and Graphs of Complexes}\label{sub:crns}

Chemical species are interconverted via chemical reactions $\rct\in\setrct=\{\pm1, \pm2,\dots,\pm\nrct\}$ characterized by the equation
\begin{equation}\small
\sum_{\spex}\stoc{}{\spex}\,\chemx  +
\sum_{\spey}\stoc{}{\spey}\,\chemy
\ch{ <=>[$\rct$][$-\rct$]}
\sum_{\spex}\stoc{-}{\spex}\,\chemx  +
\sum_{\spey}\stoc{-}{\spey}\,\chemy\,.
\label{eq:cr}
\end{equation}
Here, $\chem$ is the chemical symbol of the species $\spe$,
while $\stoc{}{\spe}$ (resp. $\stoc{-}{\spe}$) is its stoichiometric coefficient in reaction $\rct$ (resp. $-\rct$).
The set of species $\setspe$ might include non-reacting species (labeled for instance $\nr$).
All reactions are assumed to be reversible:
for every (forward) reaction $\rct\in\setrct$, reaction $-\rct\in\setrct$ and denotes its backward counterpart.

On the one hand, chemical reactions~\eqref{eq:cr} can be represented as a hypergraph, known as chemical reaction network (CRN), by mapping species into nodes and reactions into edges~\cite{Klamt2009}.
Its topology is encoded in the $\nspe\times\nrct$ stoichiometric matrix $\matS$ 
whose entries (for $\rct>0$),
\begin{equation}
\matSi{}=\stoc{-}{\spe} - \stoc{}{\spe}\,,
\label{eq:matS}
\end{equation}
specify the net variation of molecule number of species $\spe$ in reaction $\rct$. 
By definition, $\matSi{-}= -\matSi{}$ and $\matSs{}{\nr} =0$ $\forall\rct$. 
 
On the other hand, the set of chemical reactions~\eqref{eq:cr} can also define the so-called graph of complexes. 
The $\ncom$ complexes (identified by the labels $\com\in\setcom$) are aggregates of internal species acting as reactants of a reaction,
\begin{equation}
\comp{(\rct)} = \sum_{\spex}\stoc{}{\spex}\,\chemx\,,
\end{equation}
and the graph of complexes is obtained using the complexes as nodes and reactions as edges:
\begin{equation}
\comp{(\rct)}
\ch{ <=>[ $\rct$ ][ $-\rct$ ]}
\comp{(-\rct)} \,.
\label{eq:cr2}
\end{equation}
Note that the same complex can be involved in different reactions.
For instance, if reactions have no internal species among their reactants, 
they share the same complex 
(usually represented by the symbol of empty set, i.e., $\emptyset$, 
as done in the example in Subs.~\ref{subs:brusselator}).
The topology of the graph of complexes is encoded in the $\ncom\times\nrct$ incident matrix $\matI$ whose entries (for $\rct>0$),
\begin{equation}
\matIe = \dk_{\com,\com(-\rct)} - \dk_{\com, \com(\rct)},
\end{equation}
specify whether the complex $\com$ is a product (positive value) or a reagent (negative value) of reaction $\rct$ 
(with $\dk_{\com,\com'}$ being the Kronecker delta).

The graph of complexes allows us to divide the set of reactions $\setrct$ into $2\npth$ disjoint \textit{\pathways}
identified by the labels $\pth\equiv(\com,\com')\in\setpth$, 
that collect the reactions interconverting the complex $\comp{}$ into $\comp{'}$.
Namely,
\begin{equation}
\setrct = \bigcup_{\pth}\setrctpth{}\,,
\end{equation}
with 
\begin{equation}
\setrctpth{}\equiv \big\{\rct\in \setrct
\text{ such that }
\comp{(\rct)} = \comp{}
\text{ and }
\comp{(-\rct)} = \comp{'}\big\}\,.
\end{equation}
Note that, if $\rct\in\setrctpth{}$, then $-\rct\in\setrctpth{-}$ with $-\pth\equiv(\com', \com)$.
By definition, all reactions $\rct\in\setrctpth{}$ have
the same stoichiometric coefficient (resp. matrix) for each internal species $\stocx{}$ (resp. $\matSx{}$), 
labeled $\stocxe{}$ (resp. $\matSxe{}$) hereafter.
On the other hand, there might be two reactions $\rct$ and $\rct'$ in $\setrctpth{}$ such that $\matSy{}\neq\matSyy{}$.

We conclude this subsection by introducing a topological quantity, called deficiency, 
which essentially quantifies the number of ``hidden'' cycles of CRNs 
(i.e., cycles that have not a graphical representation in the graph of complexes)
and is known to underlie the emergence of complex behavior in ideal CRNs~\cite{Feinberg1987, Craciun2006, Lubensky2010, Anderson2015, Polettini2015, Marehalli2023}.
A cycle is formally defined as a right-null eigenvector $\cyclev = (\dots,\cycle,\dots)$ 
of the substoichiometric matrix for the internal species $\matSX$:
\begin{equation}\label{eq:phi}
\sum_{\rct>0}\matSx{}\cycle{}=0 .
\end{equation}
It thus represents a sequence of reactions (by specifying the number of times each reaction occurs) 
that upon completion leaves the molecule number of the internal species unchanged.
Some cycles can be represented as loops in the graph of complexes.
Indeed, the substoichiometric matrix for the internal species $\matSX$ can be expressed
in terms of the incidence matrix $\matI$ using the composition matrix $\matC$ as
\begin{equation}
\matSx{} = \sum_\com \matCe{}\, \matIe\,,
\label{eq:decompositionSX}
\end{equation}
where 
\begin{equation}
\matCe{(\rct)} = \stoc{}{\spex}
\end{equation}
specifies the stoichiometric coefficient of each internal species $\spex$ in each complex $\com$.
Hence, from Eqs.~\eqref{eq:phi} and~\eqref{eq:decompositionSX}, it follows that any right-null eigenvector of the incidence matrix $\matI$, defining a loop in the graph of complexes,
corresponds to a cycle.
However, the converse is not true and this difference is encoded in the deficiency
\begin{equation}
\deff \equiv \text{dim ker}( \matSX ) - \text{dim ker} (\matI)\,,
\label{eq:dim}
\end{equation}
where $\text{dim ker}(\bullet)$ returns the dimension of the kernel of a matrix, i.e., 
the number of linearly-independent right-null eigenvectors.

\subsection{Example}\label{subs:example0a}
\begin{figure*}
    \centering
    \includegraphics[width=0.98\textwidth]{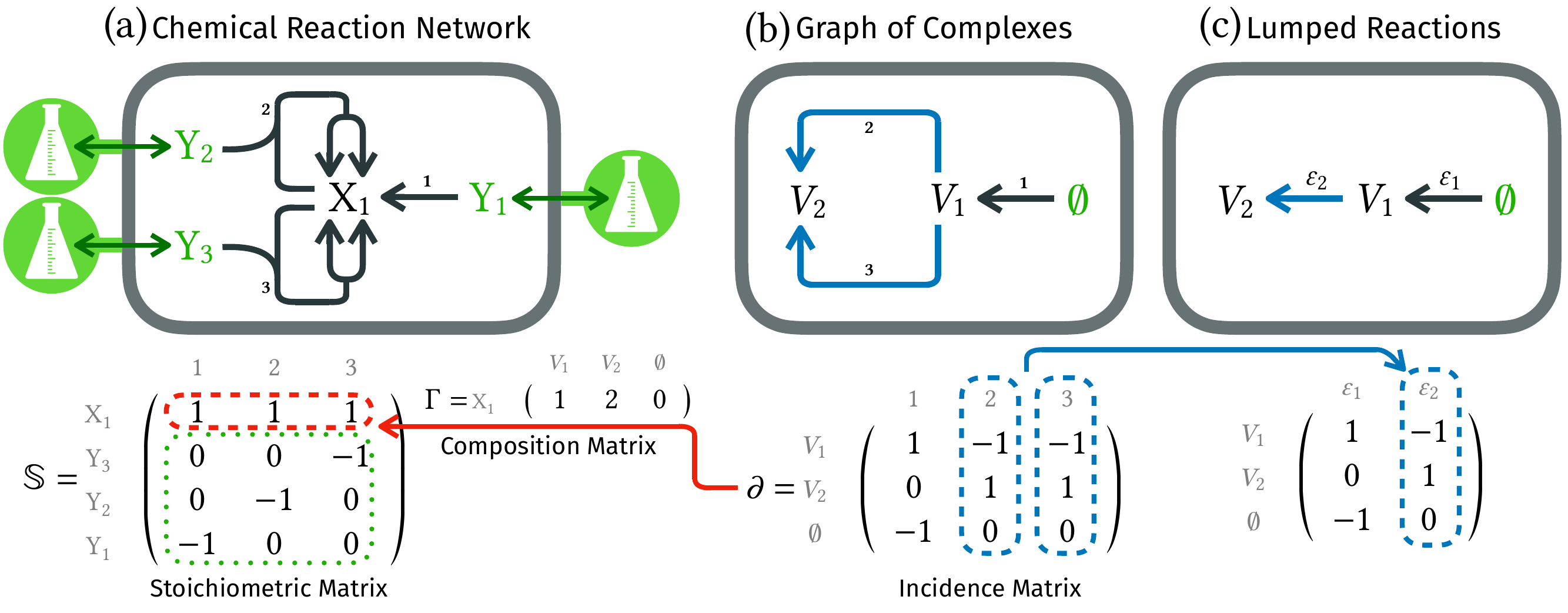}
    \caption{Different representations of the chemical reactions in Eq.~\eqref{eq:ex0:crn}.
    (a) CRN representation in terms of a hypergraph where 
    the species $\{\ch{X_1}, \ch{Y_1}, \ch{Y_2}, \ch{Y_3}\}$ are mapped into nodes, 
    while the reactions $\rct\in\{1,2,3\}$ are mapped into edges. 
    The green double-headed arrows represent the exchange processes of the chemostatted species.
    The topological properties of the CRN are encoded in its stoichiometric matrix $\matS$,
    where the red dashed (resp. green dotted) line highlights 
    the substoichiometric matrix for the internal (resp. chemostatted) species $\matSX$ (resp. $\matSY$).
    (b) Graph of complexes representation where 
    the complexes $\emptyset = \ch{Y_1}$, $V_1 = \ch{X_1}$, and $V_2 = \ch{2 X_1}$ are mapped into nodes,
    while the reactions $\rct\in\{1,2,3\}$ are mapped into edges, 
    and the exchange processes are not represented.
     The topological properties of the graph of complexes are encoded in its incidence matrix $\matI$.
     The substoichiometric matrix for the internal species $\matSX$ is related to the incidence matrix $\matI$ 
     by the composition matrix $\matC$.
     The reactions $\rct = 2$ and $\rct = 3$ (highlighted by the blue dashed line) 
     interconvert the same pair of complexes, i.e., $V_1$ and $V_2$, 
     and thus correspond to a single \pathway.
     (c) Simplified representation of the graph of complexes where 
     the \pathways{ } $\pth_1 = (\emptyset, V_1)$ and $\pth_2 = (V_1, V_2)$ are mapped into edges instead of the reactions. 
     Note that all reactions, as well all \pathways, are assumed to be reversible 
     even though only the forward reactions and \pathways{} are represented. 
     \label{fig:explanation0a}}
\end{figure*}

Consider a system composed of 
one internal species $\{\ch{X_1}\}$
and three chemostatted species $\{\ch{Y_1}, \ch{Y_2}, \ch{Y_3}\}$
which are interconverted by the following chemical reactions:
\begin{equation}
\begin{split}
\ch{Y_1} &\ch{<=>[$1$][$-1$]}\ch{X_1} \,, \\
\ch{X_1} + \ch{Y_2} &\ch{<=>[$2$][$-2$]}2\ch{X_1} \,, \\
\ch{X_1} + \ch{Y_3} &\ch{<=>[$3$][$-3$]}2\ch{X_1} \,.
\end{split}
\label{eq:ex0:crn}
\end{equation}
The first pair of reactions $\{\pm1\}$ represents the direct interconversion of $\ch{Y_1}$ into $\ch{X_1}$,
while the second (resp. third) pair of reactions $\{\pm2\}$ (resp. $\{\pm3\}$) represents the interconversion of $\ch{Y_2}$ (resp. $\ch{Y_3}$) into $\ch{X_1}$ promoted by $\ch{X_1}$ in an autocatalytic way.
In Fig.~\ref{fig:explanation0a}, we illustrate the CRN and graph of complexes representations introduced in Subs.~\ref{sub:crns}
for the chemical reactions in Eq.~\eqref{eq:ex0:crn}.

Note that the substoichiometric matrix $\matSX$ in Fig.~\ref{fig:explanation0a} admits two (linearly-independent) cycles, namely,
\begin{equation}
\cyclev=
\kbordermatrix{
   & \cr
   \color{g}\ch{1} 	 	     & 0      \cr
   \color{g}\ch{2}  		     & 1      \cr
   \color{g}\ch{3}  	  	     &-1      \cr
 }
\quad\text{ and }\quad
\cyclev'=
\kbordermatrix{
   & \cr
   \color{g}\ch{1} 	 	     & 1      \cr
   \color{g}\ch{2}  		     & 0      \cr
   \color{g}\ch{3}  	  	     &-1      \cr
 }\,.
\end{equation}
The former is also a right-null eigenvector of the incidence matrix
corresponding to the loop 
\begin{equation}
V_1 \ch{->[$2$][$$]}V_2 \ch{->[$-3$][$$]}V_1\,,
\end{equation}
while the latter is not.
Hence, $\deff = 1$.

\section{Dynamics\label{sec:dyn}}
The abundances of the chemical species $\spe\in\setspe$ in every point of space $\pos\in V$
are specified by the concentration fields $\conv = (\dots,\conc,\dots)$ following the RD equation:
\begin{equation}
\pt \conc = 
-\ds\cdot\dcurr
+\sum_{\rct>0}\matSi{}\,\rcurr{}
+ \ecurr\,,
\label{eq:rdeq}
\end{equation}
where $\ecurr$ is the exchange current of species $\spe$ with the corresponding chemostat (and thus $\ecurrx = 0$ $\forall \pos$);
$\rcurr{}$ is the net current of reaction $\rct>0$ 
given by difference between the forward $\rflux{}\geq0$ and backward $\rflux{-}\geq0$ reaction flux,
\begin{equation}
\rcurr{} = \rflux{} - \rflux{-}\,;
\end{equation}
$\dcurr$ is the diffusion current of species $\spe$. 
For sake of compactness, we omit the time-dependence of any function (see for instance the concentration fields in Eq.~\eqref{eq:rdeq}).

\remark
Note that we implicitly assumed that the concentration fields of the chemical species are large enough 
that their dynamics becomes deterministic~\cite{falasco2024macroscopic}.

\subsection{Steady State\label{sub:ss}}

The steady-state solutions of the RD equation~\eqref{eq:rdeq}, i.e., $\pt\conv = 0$, if they exist, 
can be classified as either equilibrium or nonequilibrium steady states.
Equilibrium steady states $\conveq$ satisfy 
\begin{equation}
\rcurreq{} = 0  \text{ }\text{ }\text{ }\text{ and }\text{ }\text{ }\text{ } \dcurreq = 0
\label{eq:eq}
\end{equation}
for all $\rct\in\setrct$, $\spe\in\setspe$, and $\pos\in V$.
When the RD dynamics admits well-defined equilibrium steady states, RD systems are said to be \textit{detailed balanced}.
Nonequilibrium steady states $\convss$ are characterized by 
\begin{equation}
 \rcurrss{} \neq0 \text{ }\text{ }\text{ }\text{ and/or }\text{ }\text{ }\text{ }\dcurrss \neq 0  \,,
\label{eq:ss}
\end{equation}
for some $\rct\in\setrct$ and/or $\spe\in\setspe$.

If the steady-state concentration fields are constant for every $\pos\in V$, they are said to be \textit{homogeneous} and represented by 
$\conveqh$ and $\convssh$ for equilibrium and nonequilibrium steady states, respectively.
In general, steady-state concentration fields are $\pos$-dependent and are said to be \textit{inhomogeneous}.
Also equilibrium steady states can be inhomogeneous when chemical species interact 
(as predicted by the Cahn-Hilliard~\cite{cahn1958free} and Flory-Huggins~\cite{Huggins1941,Flory1942} theories).

\subsection{Conservation Laws\label{sub:claw}}

The conservation laws~\cite{Polettini2014,Rao2016,Falasco2018a} of the RD dynamics~\eqref{eq:rdeq} (labeled $\ilaw\in\setilaw$)
are linearly-independent left-null eigenvectors $\clawv = (\dots,\claw,\dots)$ of the stoichiometric matrix, 
namely,
\begin{equation}
\sum_{\spe}\claw\, \matSi{} = 0 
\quad
\forall \rct \,.
\label{eq:claw}
\end{equation}
They physically identify fragments of (or entire) molecules, called moieties, 
that remain unaltered by the chemical reactions and the diffusion processes:
they are transferred from one species to another via the chemical reactions and 
from one point in space to another via the diffusion processes. 
Indeed, the moiety abundances, given by
\begin{equation}
\cqua = \sum_{\spe}\claw \int_V\dd\pos\,\conc\,,
\label{eq:cqua}
\end{equation}
are conserved by the RD dynamics~\eqref{eq:rdeq} in closed RD systems 
(i.e., when $\ecurr = 0$ $\forall\spe$):
\begin{equation}
\begin{split}
\dt\cqua = &
\sum_{\rct>0} \underbrace{\sum_{\spe}\claw\,\matSi{}}_{=0} \int_V\dd\pos\,\rcurr{} 
\\
&-\sum_{\spe}\claw \underbrace{\int_V\dd\pos\, \ds\cdot\dcurr}_{=0} = 0\,,
\end{split}
\end{equation}
where we used the definition of conservation law in Eq.~\eqref{eq:claw} 
and the fact that the volume $V$ has impermeable walls, i.e., $\int_V\dd\pos\, \ds\cdot\dcurr=0$. 
Note that there is always at least the conservation law $\clawmv = (\dots, \mi,\dots)$
(with $\mi$ being the molecular mass of species $\spe$)
since the total mass is conserved by the chemical reactions. 
Furthermore, the set of conservation laws $\{\clawv\}$ is not unique: 
a linear combination of conservation laws is still a conservation law. 
Different sets (i.e., different representations) identify different moieties.

In open RD systems, when the $\setspey$ species are chemostatted, 
conservation laws are split into unbroken and broken conservation laws~\cite{Rao2016,Avanzini2021}, 
labeled $\ilawu\in\setilawu$ and $\ilawb\in\setilawb$, respectively
{(this splitting will be crucial in Subs.~\ref{sub:slaw} to identify the energetic cost of breaking the detailed balance condition and maintaining RD systems out of equilibrium).
}
{The unbroken conservation laws}
are the largest subset of conservation laws that can be written with null entries for the chemostatted species,
i.e., $\law{\ilawu}{\spey} = 0$ $\forall \spey$ and $\forall\ilawu$, and hence $\sum_{\spex} \clawux \matSx{} = 0 $.
They identify moieties that are carried by the internal species only and, consequently, are not exchanged with the chemostats.
The corresponding abundances are still conserved,
\begin{equation}
\dt\cquau = 
\sum_{\spey} \underbrace{\law{\ilawu}{\spey}}_{=0} \int_V\dd\pos\,\ecurry = 0\,,
\label{eq:dtcquaub}
\end{equation}
despite the RD system being open.
{The broken conservation laws}
are the other conservation laws: $\setilawb = \setilaw \setminus \setilawu$.
They identify moieties that are carried by the chemostatted species too and, consequently, are exchanged with the chemostats.
The corresponding abundances are, in general, not conserved:
\begin{equation}
\dt\cquab = 
\sum_{\spey} \underbrace{\law{\ilawb}{\spey}}_{\neq 0} \int_V\dd\pos\,\ecurry \neq 0\,.
\label{eq:dt_brokencq}
\end{equation}
Note that open RD systems have always at least one broken conservation law 
since the total mass is not conserved by the exchange processes
and, consequently, $\clawmv$ is broken. 

In Subs.~\ref{sub:slaw}, we will use the unbroken conservation laws
to quantify the free energy exchanged between RD systems and chemostats.
To this aim, it is important to recognize that
chemostatting a species does not always break a conservation law~\cite{Falasco2018a, Avanzini2021}.
Hence, the set of chemostatted species~$\setspey$ (and the set of the corresponding chemostats) can be split 
into two disjoint subsets.

\begin{figure}
    \centering
    \includegraphics[width=0.49\textwidth]{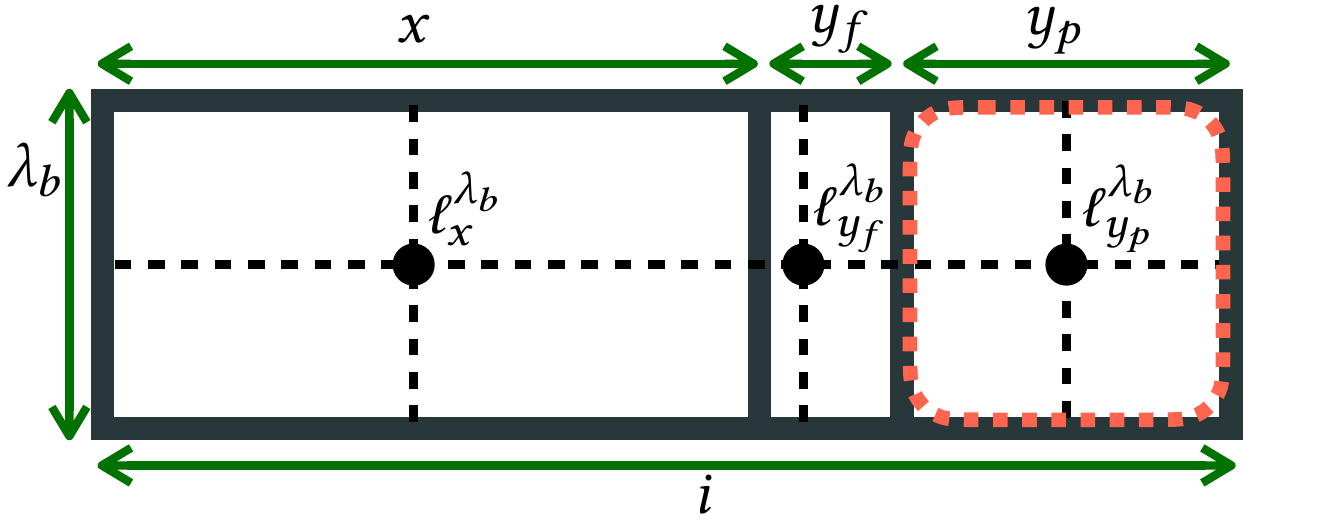}
    \caption{
    Representation of the matrix with entries $\{\clawb\}$ where $\ilawb$ is the row index and $\spe$ is the column index
    (namely, the matrix constructed using the broken conservation laws $\{\clawvb\}$ as rows).
    The matrix column (resp. the entries) can be further specialized into those corresponding to 
    the internal species $\spex$ (resp. $\clawbx$), 
    force chemostatted species $\speyf$ (resp. $\clawbf$), 
    and potential chemostatted species $\speyp$ (resp. $\clawbp$).
    The submatrix with entries $\{\clawbp\}$ (red dotted line) is square and invertible~\cite{Rao2016,postmodernthermo2023} 
    and we label $\{\iclawbp\}$ the entries of the inverse matrix 
    where $\speyp$ is the row index and $\ilawb$ is the column index.
     \label{fig:explanation0c}}
\end{figure}

The potential species (labeled $\speyp\in\setspeyp$) are the smallest subset of chemostatted species such that 
all conservation laws $\{\clawvb\}$ with $\ilawb\in\setilawb$ are broken.
Crucially, there is always a representation of the broken conservation laws such that
each corresponding moiety is exchanged with only one potential chemostat.
Indeed, the matrix 
(see the graphical representation in Fig.~\ref{fig:explanation0c}) 
whose entries are $\{\clawbp\}$ (with $\ilawb\in\setilawb$ and $\speyp\in\setspeyp$) 
is square and invertible~\cite{Rao2016,postmodernthermo2023} and,
by labeling $\{\iclawbp\}$ the entries of the inverse matrix, 
the broken conservation laws can be represented in terms of the following linear combination as
\begin{equation}
\sum_{\ilawb} \iclawbp \, \clawvb\,.
\label{eq:rep_brokencl}
\end{equation}
The abundance of each moiety in this representation
\begin{equation}
\cmoi = \sum_{\spe} \sum_{\ilawb} \iclawbp \,  \clawb \int_V\dd\pos\,\conc\,,
\label{eq:moiety} 
\end{equation}
where $\spe$ runs over all specie $\setspe$,
changes (according to the RD equation~\eqref{eq:rdeq}) only because of the exchanges with one specific potential chemostat:
\begin{equation}\small
\dt \cmoi 
= \sum_{ \speypp}\underbrace{\sum_{\ilawb} \iclawbp \, \clawbpp}_{=\dk_{\speyp,\speypp}} \int_V\dd\pos\,\ecurrypp
= \int_V\dd\pos\,\ecurryp
\label{eq:dt_moieity_yp}
\end{equation}
if all chemostatted species are potential species, i.e., $\setspey = \setspeyp$.
This allows us to label each moiety in Eq.~\eqref{eq:moiety} using the label of the chemostat it is exchanged with, i.e., $\speyp$, 
rather than the label of the corresponding broken conservation law, i.e., $\ilawb$.
Note that Eq.~\eqref{eq:dt_moieity_yp} specializes Eq.~\eqref{eq:dt_brokencq} 
for open RD systems with only potential chemostatted species 
and with the conservation laws expressed according to the representation in Eq.~\eqref{eq:rep_brokencl}.
We emphasize that the representation in Eq.~\eqref{eq:rep_brokencl} will be the one used to
quantify the free energy exchanged between RD systems and chemostats 
by entering the definition of the thermodynamic potential in Eq.~\eqref{eq:sgfree}.

The force species ($\speyf\in\setspeyf$) are the other chemostatted species: $\setspeyf = \setspey \setminus \setspeyp$.
The corresponding chemostats exchange moieties that are already exchanged with the potential chemostats.
Indeed, in general, the abundance of each moiety in the representation given in Eq.~\eqref{eq:rep_brokencl} 
changes according to
\begin{equation}
\dt \cmoi 
= \int_V\dd\pos\,\ecurryp + \sum_{ \speyf} \sum_{\ilawb} \iclawbp \, \clawbf \int_V\dd\pos\,\ecurryf \,,
\label{eq:dt_moieity}
\end{equation}
using Eqs.~\eqref{eq:rdeq} and~\eqref{eq:moiety} and the splitting $\setspey = \setspeyp \bigcup \setspeyf$.
Hence, chemostatting the force species (after the potential species were already chemostatted)
does not break any new conservation law, 
but creates a flux of the same moiety between different chemostats.
The energetic cost of sustaining these fluxes will enter the definition of the nonconsevative work in Eq.~\eqref{eq:ncwrk}.

\subsection{Example}\label{subs:example0b}
\begin{figure}
    \centering
    \includegraphics[width=0.49\textwidth]{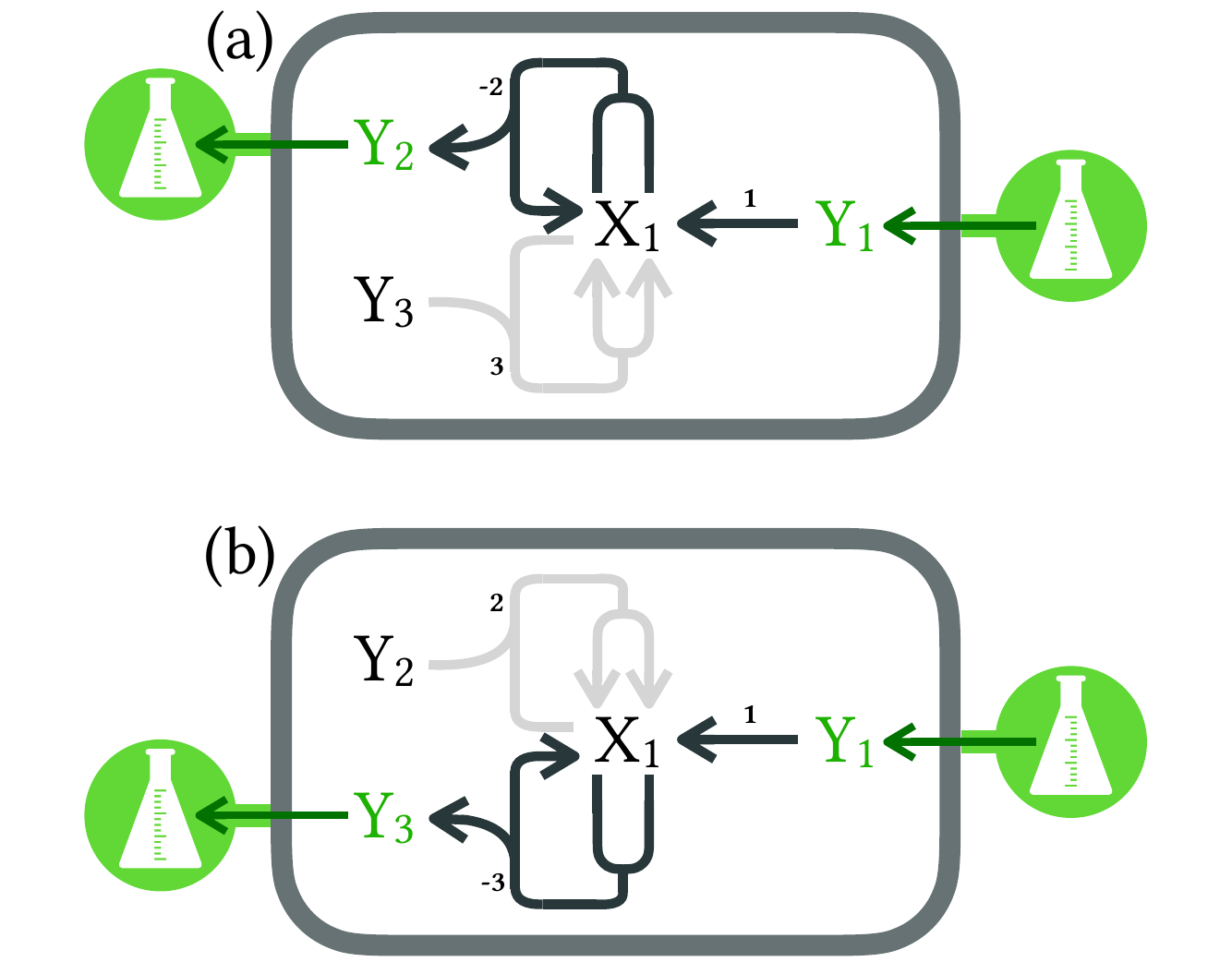}
    \caption{Illustration of the sequences of reactions transferring the same moiety from a potential chemostat to a force chemostat 
    (for a RD system with the chemical species and reactions introduced in Subs.~\ref{subs:example0a}).
    (a) When only the species \ch{Y_1} and \ch{Y_2} are chemostatted, 
    the moiety can be injected into the RD system by a chemostat as the species $\ch{Y_1}$,
    transferred via the sequence of reaction $\{1, -2\}$ to the species $\ch{Y_2}$,
    and finally extracted from the RD system by a different chemostat as the species $\ch{Y_2}$.
    (b) When only the species \ch{Y_1} and \ch{Y_3} are chemostatted, 
    the moiety can be injected into the RD system by a chemostat as the species $\ch{Y_1}$,
    transferred via the sequence of reaction $\{1, -3\}$ to the species $\ch{Y_3}$,
    and finally extracted from the RD system by a different chemostat as the species $\ch{Y_3}$.
    Note that when all three species $\{\ch{Y_1},\ch{Y_2},\ch{Y_3}\}$ are chemostatted,
    the same moiety can be transferred between each pair of chemostats.
    \label{fig:explanation0b}}
\end{figure}

A RD system with the chemical species and reactions introduced in Subs.~\ref{subs:example0a} 
admits only one conservation law (see the stoichiometric matrix in Fig.~\ref{fig:explanation0a}), i.e.,
\begin{equation}
\clawve=
\kbordermatrix{
   & \cr
   \color{g}\ch{X_1} 	 	     & 1      \cr
   \color{g}\ch{Y_3}  		     & 1      \cr
   \color{g}\ch{Y_2}  	  	     & 1      \cr
   \color{g}\ch{Y_1}   		     & 1      \cr
 }\,.
\label{eq:ex:cl}
\end{equation}
This physically means that all species $\{\ch{X_1}, \ch{Y_1}, \ch{Y_2}, \ch{Y_3} \}$ are made of the same atoms,
namely, the same moiety.
Therefore, 
$\cquae = \int_V\dd\pos\, (
c_{\ch{X_1}}(\pos) + 
c_{\ch{Y_1}}(\pos) + 
c_{\ch{Y_2}}(\pos) +
c_{\ch{Y_3}}(\pos))$
is conserved if no species are chemostatted. 
By chemostatting just one of the species $\{\ch{Y_1}, \ch{Y_2}, \ch{Y_3} \}$, the conservation law is broken.
Hence, any of the species $\{\ch{Y_1}, \ch{Y_2}, \ch{Y_3} \}$ can be chosen as a potential species.
We choose \ch{Y_1}. 
By chemostatting another species, like for instance \ch{Y_2}, no conservation law is broken.
Hence, \ch{Y_2} is a force species and
the same moiety is exchanged between 2 different chemostats as illustrated in Fig.~\ref{fig:explanation0b}.
Note that this implies that the matrix with entries $\{\clawbp\}$ has just one entry corresponding to $\clawbpex = 1$ and that the conservation law in Eq.~\eqref{eq:ex:cl} is already written in the representation given in Eq.~\eqref{eq:rep_brokencl}.

\section{Thermodynamics\label{sec:thermo}}
We now develop a nonequilibrium thermodynamic theory for open RD systems of interacting chemical species undergoing the RD dynamics~\eqref{eq:rdeq} by combining the thermodynamics of
noninteracting (i.e., ideal) RD systems~\cite{Falasco2018a, Avanzini2019a}
and interacting (i.e., non-ideal) CRNs~\cite{Avanzini2021}.

\subsection{Nonequilibrium Free Energy}

Nonequilibrium thermodynamics presumes that all degrees of freedom other than concentration fields are equilibrated (e.g., the temperature $T$ and the volume $V$ are fixed, while  the fields responsible for the interactions, e.g., electrostatic fields, relax instantaneously to mean-field values depending on the concentration fields only).
In this framework, the nonequilibrium Helmholtz free energy $\free$ of RD systems can be specified by its equilibrium form, but as a functional of the nonequilibrium concentration fields.
Furthermore, $\free$ can always be written as the sum of an \textit{ideal} and a \textit{non-ideal} contribution:
\begin{equation}
\free= \freeid + \freeni\,.
\label{eq:free}
\end{equation}
Here, the ideal term $\freeid$ is the ideal solution Helmholtz free energy,
while the non-ideal term $\freeni$ accounts for the interactions between the chemical species.

The free energy contribution of each chemical species is specified by the chemical potential
\begin{equation}
\cp = \Fd{\free}= \cpid +RT \ln \act\,,
\label{eq:cp}
\end{equation}
where $\cpid$ is the ideal chemical potential
\begin{equation}
\cpid
= \stcp + RT \ln{\conc}\,,
\label{eq:cp_id}
\end{equation}
with $\stcp$ the standard chemical potential, $R$ the gas constant,
and $\act$ the activity coefficient accounting for the interactions,
\begin{equation}
RT \ln \act = \Fd{\freeni} \,.
\label{eq:act}
\end{equation}
Note that the ideal free energy $\freeid$ reads~\cite{Falasco2018a, Avanzini2019a} 
\begin{equation}\label{eq:f_ideal}
 \freeid = \sum_{\spe}\int_V\dd\pos\, {\big(\cpid-RT\big)\conc} \,.
\end{equation}

\subsection{Thermodynamic Consistency of the RD Dynamics\label{subs:thermo_cost}} 

We start by introducing the entropy production rate
\begin{equation}
\epr \equiv \eprrct + \eprdff\,,
\label{eq:epr0}
\end{equation}
that quantifies the dissipation resulting from the changes in chemical potentials caused by the two types of processes occurring in the system, namely, reactions and diffusion:
\begin{subequations}\label{eq:epr}
\begin{align}
T\eprrct &\equiv -\sum_{\spe} \sum_{\rct>0}\int_V\dd\pos\, \cp \matSi{}\rcurr{}\label{eq:eprrct1}\,,\\
T\eprdff &\equiv - \sum_{\spe}\int_V\dd\pos\, \ds\cp\cdot\dcurr \,. \label{eq:eprdff1}
\end{align}%
\end{subequations} 
The central assumption ensuring thermodynamic consistency of the RD dynamics is that the reaction fluxes $\rflux{\pm}$ satisfy the local detailed balance condition:
\begin{equation}\small
RT\ln\frac{\rflux{}}{\rflux{-}} = 
-\Big(
\sum_{\spex}\cpx\matSx{}
+\sum_{\spey}\cpy\matSy{}
\Big)\,,
\label{eq:ldb}
\end{equation}
{resulting from the macroscopic limit~\cite{falasco2024macroscopic}
of the local detailed balance condition of stochastic thermodynamics~\cite{Pigolotti:ST} 
when applied to CRNs~\cite{Rao2018b,postmodernthermo2023}}.
We further assume that the diffusion current $\dcurr$ of each species is treated within linear regime and expressed as
\begin{equation}
\dcurr = -\sum_{\spee}\matOij\ds\cpg{\spee}\,, 
\label{eq:dcurr}
\end{equation}
where $\{\matOij\}$ are the entries of the positive-definite Onsager matrix $\matO$~\cite{Groot1984}.
Using these two assumptions in Eqs.~\eqref{eq:ldb} and~\eqref{eq:dcurr}, 
we see that the two entropy production rates in Eq.~\eqref{eq:epr} can be rewritten in a manifestly non-negative form
\begin{subequations}\label{eq:epr2}\small
\begin{align}
T\eprrct &= \int_V\dd\pos\underbrace{ RT\sum_{\rct>0} \rcurr{}\ln\frac{\rflux{}}{\rflux{-}}}_{\equiv T\deprrct\geq 0}\geq0\,,\label{eq:eprrct2}\\
T\eprdff &=  \int_V\dd\pos\underbrace{\sum_{\spe,\spee} \ds\cp \cdot\matOij  \ds\cpg{\spee}}_{\equiv T\deprdff\geq 0} \geq0\,,\label{eq:eprdff2}
\end{align}%
\end{subequations}
where the equality sign is achieved only at equilibrium~\eqref{eq:eq}.

The change in free energy~\eqref{eq:free} in closed RD systems, 
using Eq.~\eqref{eq:rdeq} (when $\ecurr = 0$ $\forall\spe$) 
and integrating by parts, 
gives the second law of thermodynamics for closed RD systems
\begin{equation}
\dt\free = - T \epr \leq 0\, .
\end{equation}
With the additional assumption that the free energy in Eq.~\eqref{eq:free} is lower bounded, 
$\free$ becomes a thermodynamic potential acting as a Lyapunov function of the RD dynamics~\eqref{eq:rdeq}.
This ensures  thermodynamic consistency, 
namely, closed RD systems always relax towards an equilibrium steady state.

\remark
In Eqs.~\eqref{eq:eprrct2} and~\eqref{eq:eprdff2} 
we introduced the densities of reaction $\deprrct$ and diffusion $\deprdff$ entropy production rate, respectively.
They are non-negative for every $\pos\in V$ and vanish only at equilibrium~\eqref{eq:eq}.
The density of reaction entropy production rate can be also split into 
the non-negative contribution of each reaction $\deprrctr\geq 0$:
$\deprrct = \sum_{\rct > 0} \deprrctr$.

\remark
The expressions of entropy production rates in Eqs.~\eqref{eq:eprrct2} and~\eqref{eq:eprdff2} 
together with the local detailed balance condition~\eqref{eq:ldb} and 
$\matO$ being a \cmark{positive-definite} matrix
imply that the thermodynamic equilibrium conditions, 
$\eprrcteq = 0$ and $\eprdffeq = 0$, are equivalent to
\begin{subequations}
\begin{align}
&\sum_{\spe} \cpeq \matSi{} = 0\,, \label{eq:teqrct}\\
&\ds\cpeq  = 0\,,\label{eq:teqdff}
\end{align}\label{eq:teq}%
\end{subequations}
for all $\pos\in V$, for all $\rct\in\setrct$, and for all $\spe\in\setspe$.
The equilibrium conditions in Eq.~\eqref{eq:teq} can also be derived by minimizing the free energy~\eqref{eq:free}
as done in App.~\ref{app:free_var}.

\subsection{Chemostats\label{sub:chemo}}
Thermodynamically,
chemostats control the chemical potentials $\cpy$ of the chemostatted species in every point of space $\pos\in V$,
which can thus be treated as (in general, $\pos$- and time-dependent) control parameters, i.e,  $\{\cpry\}$.
In ideal RD systems, where the chemical potentials are given in Eq.~\eqref{eq:cp_id}, 
chemostats control the concentration fields of the chemostatted species too: 
$\cony = \exp((\cpry - \stcpy)/RT)$. 
The exchange current $\ecurry$ in Eq.~\eqref{eq:rdeq} represents the physical mechanism implementing this control.
In non-ideal RD systems, where the chemical potentials are given in Eq.~\eqref{eq:cp},
controlling the  chemical potentials of the chemostatted species do not univocally determine their concentrations 
because of the interactions~\cite{Avanzini2021}.

\subsection{Second Law for Open RD Systems \label{sub:slaw}}

Using the expression for the entropy production rate in Eq.~\eqref{eq:epr0},
{together with the splitting of the conservation laws in broken and unbroken 
as well as the splitting of the chemostatted species into potential and force species},
the second law of thermodynamics for open RD systems can now be rewritten as 
\begin{equation}
T\epr = -\dt \sgfree + \ncwrk + \drwrk \geq 0 \,,
\label{eq:slaw}
\end{equation}
This can be checked by direct substitution 
of the different terms on the right-hand side that we now define 
together with the RD equation~\eqref{eq:rdeq}.

The semigrand Helmholtz free energy $\sgfree $ is the proper thermodynamic potential of open RD systems. 
It reads
\begin{equation}
\sgfree = \free - \sum_{\speyp} \cprypref \,\cmoi
\label{eq:sgfree}
\end{equation}
and is obtained, in analogy to equilibrium thermodynamics when passing from the canonical to the grand canonical ensembles,
from the Helmholtz free energy~\eqref{eq:free} 
by removing the energetic contribution of the matter exchanged with the chemostats.
The latter accounts for the abundances of the moieties $\{\cmoi\}$, given in Eq.~\eqref{eq:moiety},
times a reference value of chemical potentials of the potential species $\{\cprypref\}$, controlled by the corresponding chemostats.
If the potential chemostats impose different values of the chemical potentials $\{\cpryp\}$ in different $\pos\in V$, 
then the reference chemical potential $\{\cprypref\}$ can be chosen arbitrarily among these values.
This is equivalent to choosing a reference equilibrium steady state $\conveq$ satisfying $\sgfreeeq\leq\sgfree$ 
as discussed in App.~\ref{app:sgfree_var}.

The nonconservative work rate 
\begin{equation}\small
\ncwrk 
= \int_V\dd\pos\,\underbrace{\sum_{\spey}\ncforce{\spey}\ecurry}_{\equiv \dncwrk}\\
\label{eq:ncwrk}
\end{equation}
quantifies the energetic cost for sustaining fluxes of the same moiety between chemostats with different chemical potentials
by means of the nonconservative forces 
\begin{equation}
\ncforce{\spey} \equiv \cpry - \sum_{\speyp,\ilawb}\cprypref \, \iclawbp \,  \clawby\,.
\end{equation}
These forces emerge in two cases. 
When $\cpryp \neq \cprypref $ (because $\sum_{\ilawb} \iclawbp \,  \clawbpp = \dk_{\speyp, \speypp}$), the forces $\ncforce{\speyp}$ correspond to diffusive forces that can be present even in the absence of chemical reactions~\cite{Falasco2018a}.
They arise because the potential chemostats $\speyp$ impose values of the chemical potential that are different from the reference equilibrium ones $\cprypref$ (see App.~\ref{app:sgfree_var}).
When $\cpryf \neq \sum_{\speyp,\ilawb}\cprypref \, \iclawbp \,  \clawbf $, the forces $\ncforce{\speyf}$ correspond to chemical forces that can also be present in homogeneous systems~\cite{Avanzini2021}.
They arise because force chemostats $\speyf$ impose chemical potentials that are different from the one they would have if all reactions were equilibrated $\sum_{\speyp,\ilawb}\cprypref \, \iclawbp \,  \clawbf $ (see App.~\ref{app:sgfree_var}).

The driving work rate reads
\begin{equation}
\drwrk = - \sum_{\speyp} \big(\dt\cprypref\big) \,\cmoi 
\label{eq:drwrk}
\end{equation}
and quantifies the energetic cost of the time-dependent changes of the reference chemical potentials 
$\{\cprypref\}$ (with $\speyp\in\setspeyp$) defining the reference equilibrium to which the open RD system would relax 
if the nonconservative forces vanished.

Physically, the second law~\eqref{eq:slaw} expresses (and quantifies)
the different sources of free energy (on the right-hand side of Eq.~\eqref{eq:slaw}) 
that can balance dissipation (on the left-hand side of Eq.~\eqref{eq:slaw}) 
and hence maintain open RD systems out of equilibrium. 
Let us illustrate this point by considering three simple cases.

First, when the chemical potentials of the chemostatted species are independent of time and such that $\ncforce{\spey}=0$ 
(which also implies homogeneous chemostatting), 
both work contributions vanish and the second law~\eqref{eq:slaw} reduces to $\dt\sgfree = -T\epr\leq 0$.
This, together with $\sgfree$ being lower bounded
(as shown in App.~\ref{app:sgfree_var}),
implies that the RD system relaxes to equilibrium while minimizing $\sgfree$ which serves as a Lyapunov function.
This RD system is therefore detailed balanced despite being open.
As shown in App.~\ref{app:sgfree_var}, the specific equilibrium steady state $\conveq$ to which the RD system relaxes is defined by the chemical potentials $\{\cprypref\}$ and the abundances $\{\cquau\}$.
{If self-organized structures emerged, i.e., the concentrations  $\conveq$ are not homogeneous, 
they would be sustained by a passive mechanism that does not dissipate. 
}

Second, when the chemical potentials of the chemostatted species are dependent on time but $\ncforce{\spey}=0$ still holds,
only the nonconservative work rate vanishes from the second law~\eqref{eq:slaw}, 
namely, $T\epr = - \dt\sgfree + \drwrk \geq 0$, and the RD system is still detailed balanced. 
Indeed, if the time dependence is very slow, 
the RD system will quasi-statically follow the changing equilibrium state.
For faster time dependence, the RD system will be prevented from reaching equilibrium. 

Third, when the chemical potentials of the chemostatted species are independent of time but $\ncforce{\spey} \neq 0$, 
detailed balance is broken and the second law~\eqref{eq:slaw} reduces to $T\epr = - \dt\sgfree + \ncwrk \geq 0$. If the RD system eventually reaches a nonequilibrium steady state, the second law is further reduced to $T\eprss =  \ncwrkss \geq 0$.
In this case, $\ncwrkss$ quantifies the net free energy intake by the open RD system that is dissipated to maintain the system in a nonequilibrium steady state.

\subsection{Example}
We consider 
the RD system with the chemical species and reactions introduced in Subs.~\ref{subs:example0a} 
and with the chemostats imposing $\pos$-dependent chemical potentials 
$\cpe{\ch{Y_1}}$, $\cpe{\ch{Y_2}}$, and $\cpe{\ch{Y_3}}$.
We choose $\ch{Y_1}$ as potential species (as discussed in Subs.~\ref{subs:example0b})
and one of the values of $\cpe{\ch{Y_1}}$ as reference chemical potential $\cprefe{\ch{Y_1}}$.
The corresponding nonconservative work rate (whose general expression is {given} in Eq.~\eqref{eq:ncwrk})
specializes into the sum of three contributions:
\begin{equation}
\begin{split}
\ncwrk = 
&\int_V\dd\pos\, (\cpe{\ch{Y_1}} - \cprefe{\ch{Y_1}}) \ecurre{\ch{Y_1}} 
\\
+&\int_V\dd\pos\, (\cpe{\ch{Y_2}} - \cprefe{\ch{Y_1}}) \ecurre{\ch{Y_2}} 
\\ 
+&\int_V\dd\pos\, (\cpe{\ch{Y_3}} - \cprefe{\ch{Y_1}}) \ecurre{\ch{Y_3}}\,,
\end{split}
\label{eq:ex:ncwork}
\end{equation}
where we used the broken conservation law in Eq.~\eqref{eq:ex:cl}.
The first contribution results from applying different chemical potentials to the species $\ch{Y_1}$.
The second (resp. third) contribution results from applying different chemical potentials to the same moiety 
carried by both the species $\ch{Y_1}$ and $\ch{Y_2}$ (resp. $\ch{Y_3}$) as illustrated in Fig.~\ref{fig:explanation0b}.

{Notice that the nonconservative work in Eq.~\eqref{eq:ex:ncwork} is not 
the excess work introduced in Refs.~\cite{Ross1992,Ross1993},
even though the two expressions might look similar. 
The difference between the actual chemical potential and a reference one features both expressions.
This reference chemical potential is the steady state one in the excess work, 
while it takes the equilibrium value imposed by the potential chemostats in the nonconservative work.
Furthermore, also a current multiplying the difference between chemical potentials features the expression of both works.
This current is the total time derivative of a species concentration in the excess work,
while it is the exchange current with the chemostat in the non nonconservative work.
Indeed, the excess work is not directly related to the evolution of the semigrand free energy~\eqref{eq:sgfree} 
via the second law~\eqref{eq:slaw},
but rather to the evolution of the kinetic potential in Eq.~\eqref{eq:LYcb} in homogeneous systems.
}

\section{Thermodynamic Properties of Active Self-Organization - Exact Results\label{sec:aps}}
\begin{figure}
    \centering
    \includegraphics[width=0.49\textwidth]{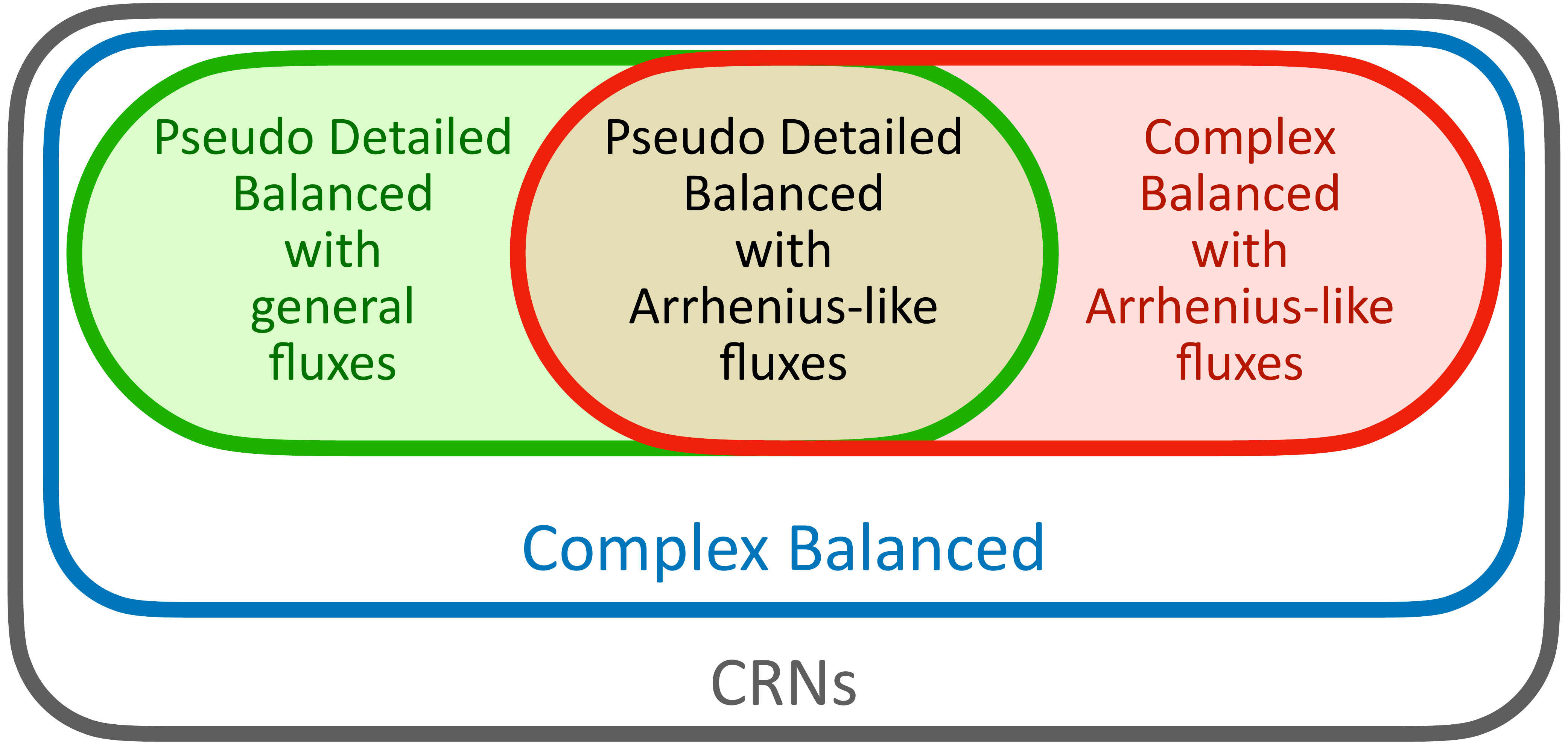}
    \caption{
    The gray box represents the set of all possible CRNs endowed with all possible reaction fluxes.
    The blue box, a subset of the gray box, represents the set of all possible complex balanced CRNs endowed 
    with all possible reaction fluxes.
    Pseudo detailed balanced CRNs are always a subset of complex balanced CRNs.
    In Sec.~\ref{sec:aps}, we consider 
    pseudo detailed balanced CRNs with reaction fluxes of general form (green oval) and 
    complex balanced CRNs with Arrhenius-like reaction fluxes (red oval). 
     \label{fig:explanation0d}}
\end{figure}

The RD dynamics might reach a steady state where the chemical species are inhomogeneous. 
We now use the thermodynamic theory introduced in Sec.~\ref{sec:thermo} 
to prove that
for two classes of CRNs, namely, pseudo detailed balanced (Subs.~\ref{sub:pdb}) and complex balanced (Subs.~\ref{sub:cb})
discussed in Fig.~\ref{fig:explanation0d}, 
stationary concentration profiles are determined by the minimization of appropriately constructed kinetic potentials 
acting as Lyapunov functions,
when the chemical potentials $\{\cpyy\}$ of the chemostatted species are homogeneous and constant in time.
Furthermore, for both classes of CRNs, diffusion processes always equilibrate, 
which implies that the chemical potentials become homogeneous at steady state.
Additionally, for pseudo-detailed CRNs only, the transient dynamics and steady-state concentration fields can be exactly simulated also in reaction-diffusion systems relaxing towards equilibrium with appropriately constructed chemical potentials.

The proofs of this result are derived in subs.~\ref{sub:pdb} and~\ref{sub:cb}
by assuming that the chemostatted species are ideal,
i.e., they do not interact with the internal species, for the sake of simplicity. 
We can thus treat both $\{\cpyy\}$ and $\{\conyy\}$ (related via Eq.~\eqref{eq:cp_id}) as control parameters.
We generalize our result to non-ideal chemostatted species in sub.~\ref{sub:pdb_cp_nonideal}.

\subsection{Pseudo Detailed Balanced CRNs\label{sub:pdb}}
We define CRNs as pseudo detailed balanced when their stoichiometric matrix for the internal species 
and \pathways{}
does not admit any right-null eigenvector.
Namely, there is no $\{\cyclee\}$ such that
\begin{equation}
\sum_{\pth>0}\matSxe{}\, \cyclee = 0\,.
\label{eq:path_balanced}
\end{equation}
We now show that the RD dynamics of pseudo detailed balanced CRNs 
is driven by the minimization of a kinetic potential and
relaxes to a steady state $\convss$ characterized by a vanishing diffusion entropy production rate, i.e., $\eprdffss = 0$,
independently of whether the concentration fields are homogeneous or inhomogeneous.

\assumption 
We consider reaction fluxes of the general form
\begin{equation}
\rflux{} = \kconst{} \rfluxx \rfluxyy\,,
\label{eq:flux_ass1}
\end{equation}
namely, the concentration-dependence of each flux $\rflux{}$ is univocally determined 
by the stoichiometric coefficients 
via the unspecified functions $\rfluxx$ and $\rfluxyy$.
The latter depends only on the concentrations of chemostatted species because they are assumed to be ideal.
Here, $\kconst{}$ is an unspecified constant parameter.
For example, mass-action fluxes~\cite{Laidler1987,Pekar2005} and Arrhenius-like fluxes satisfy Eq.~\eqref{eq:flux_ass1}.
Note that, while the latter read
\begin{equation}
\rflux{} = \AAp{}\,e^{\frac{\sum_{\spex}\cpx\stocx{}}{RT}}e^{\frac{\sum_{\spey}\cpyy\stocy{}}{RT}}\,,
\label{eq:arr_gen}
\end{equation}
with $\AAp{} = \AAp{-}$ being a constant parameter,
the fluxes in Eq.~\eqref{eq:flux_ass1} can in general account for a concentration-dependent~$\AAp{}$.

\begin{proof}
We start by using $\matSx{} = \matSxe{}$ $\forall \rct\in\setrctpth{}$ 
to rewrite the contribution due to the chemical reactions in Eq.~\eqref{eq:rdeq} for the internal species as
\begin{equation}
\sum_{\rct>0}\matSx{}\,\rcurr{} = \sum_{\pth>0}\matSxe{}\pcurr{}\,,
\label{eq:cc_path}
\end{equation}
with 
\begin{equation}
\pcurr{} = \sum_{\rct\in\setrctpth{}}\big\{\rflux{} - \rflux{-}\big\}
\label{eq:pcurr_def}
\end{equation}
being the \pathway{} $\pth$ net current.
By writing now each reaction flux $\rflux{}$ as the product between 
its symmetric part $\sr{} \equiv \sqrt{\rflux{}\rflux{-}} = \sr{-}$
and its antisymmetric part $\sqrt{\rflux{}/\rflux{-}}$ 
specified by the local detailed balance condition~\eqref{eq:ldb},
$\pcurr{}$ becomes
\begin{equation}\small
\pcurr{} = \aap{}\Big\{ e^{-\frac{\sum_{\spex}\cpx\matSxe{}}{2RT}} - \bbp{} e^{-\frac{\sum_{\spex}\cpx\matSxe{-}}{2RT}}\Big\}\,,
\label{eq:pcurr2}
\end{equation}
where we factorized
\begin{equation}
\aap{} = \sum_{\rct\in\setrctpth{}} \sr{} e^{-\frac{\sum_{\spey}\cpyy\matSy{}}{2RT}}\geq0\,,
\end{equation}
and the constant quantity
\begin{equation}
\bbp{} = \frac{\aap{-}}{\aap{}} = \frac{\sum_{\rct\in\setrctpth{}} \sr{} e^{-\frac{\sum_{\spey}\cpyy\matSy{-}}{2RT}}} {\sum_{\rct\in\setrctpth{}} \sr{} e^{-\frac{\sum_{\spey}\cpyy\matSy{}}{2RT}}}\geq 0.
\label{eq:bbp}
\end{equation}
This results from the fact that 
i) the chemical potentials $\{\cpyy\}$, and the corresponding concentrations $\{\conyy\}$, are homogeneous and constant,
and ii) and the fluxes in $\sr{}$ are univocally determined by the stoichiometric coefficients according to Eq.~\eqref{eq:flux_ass1} with $\stocx{} =\stocxe{}$ $\forall \rct\in\setrctpth{}$.

We now use the pseudo detailed balanced condition.
Since there are no right-null vectors of $\matSxe{}$, we can always introduce the constant coefficients $\{\shx\}$ such that
\begin{equation}
\ln\bbp{} = \sum_\spex \shx\, \matSxe{} /RT\,.
\label{eq:delta_def}
\end{equation}
Indeed, given the $\nspex\times\npth$ matrix $\matSSxe$ whose entries are $\{\matSxe{}\}$ 
with $\spex\in\setspex$, $\setpth\ni\pth>0$, $\rank(\matSSxe)=\npth$.
Therefore, the set of internal species can always be split into two disjoint sets $\setspexi$ and $\setspexd$ such that 
the matrix $\matSSxie$ (with entries $\{\matSxe{}\}$ with $\spex\in\setspexi$ and $\setpth\ni\pth>0$) is square and invertible.
This leads to 
\begin{subequations}
\begin{align}
\shx &= RT \sum_{\pth>0} \ln\bbp{} (\matSSxie^{-1})_{\pth,\spex} \text{ }\text{ }\text{ }\text{ for }\text{ }\text{ }\text{ } \spex\in\setspexi\,,\\
\shx &= 0 \text{ }\text{ }\text{ }\text{ for }\text{ }\text{ }\text{ } \spex\in\setspexd\,,
\end{align}
\label{eq:shift:pdb}%
\end{subequations}
and allows us to write the \pathway{} $\pth$ net current~\eqref{eq:pcurr2} as
\begin{equation}\small
\pcurr{} = \pflux{} - \pflux{-}\,.
\label{eq:pcurr3}
\end{equation}
Here, we introduced the \pathway{} $\pm\pth$ fluxes that read
\begin{equation}
\pflux{\pm} \equiv \aap{\abs} e^{\frac{\sum_\spex\shx\matSxe{\abs}}{2RT}}  e^{-\frac{\sum_{\spex}\cpxe\matSxe{\pm}}{2RT}}\,,
\end{equation}
with $\abs{-\pth} = \abs{\pth} = {\pth}>0$,
and satisfy a pseudo local detailed balance condition
\begin{equation}
RT\ln\frac{\pflux{}}{\pflux{-}} = 
- \sum_{\spex}\cpxe\matSxe{}\,,
\label{eq:ldb_pb}
\end{equation}
with the chemical potentials
\begin{equation}
\cpxe \equiv \cpx + \shx\,,
\label{eq:scp}
\end{equation}
Note that Eq.~\eqref{eq:scp} only implies a shift of the standard chemical potentials of the internal species $\{\stcpx\}$. 

Until now, we just re-wrote the \pathway{} net current $\pcurr{}$ entering the RD equation~\eqref{eq:rdeq} via Eq.~\eqref{eq:cc_path} as in Eq.~\eqref{eq:pcurr3}.
However, this rewriting allows us to show that the potential $\freepdb$, 
obtained from the free energy~\eqref{eq:free} of the RD system
after applying the shift of the chemical potentials in Eq.~\eqref{eq:scp},
decreases in time during RD dynamics~\eqref{eq:rdeq}.
Indeed, using $\dt\conyy = 0$, 
its time derivative reads
\begin{equation}
\dt\freepdb = - T\eprrcte - T\eprdff \leq 0\,,
\label{eq:lyapunov_pdb}
\end{equation}
where $\eprdff\geq 0$ is given in Eq.~\eqref{eq:eprdff2} (since $\ds \cpyy = 0$) and, using Eq.~\eqref{eq:ldb_pb},
\begin{equation}\small
T\eprrcte = RT\sum_{\pth>0}\int_V\dd\pos\, \pcurr{}\ln\frac{\pflux{}}{\pflux{-}}\geq0\,.
\label{eq:eprrcte2}
\end{equation}
This, together with the fact that 
$\freepdb$ is lower bounded (since  $\free$ is lower bounded by assumption, see Subs.~\ref{subs:thermo_cost}),
implies that the RD dynamics~\eqref{eq:rdeq} of pseudo detailed balance CRNs 
must relax towards a steady state $\convss$ satisfying
\begin{equation}
\eprrctess = 0 \text{ }\text{ }\text{ }\text{ and }\text{ }\text{ }\text{ } \eprdffss = 0\,,
\label{eq:pseudo_eq1}
\end{equation}
or equivalently
\begin{equation}
\sum_{\spex}(\cpgss{\spex} + \shx)\matSxe{} =0
\text{ }\text{ }\text{ }\text{ and }\text{ }\text{ }\text{ }
\dcurrss = 0 \,,
\label{eq:pseudo_eq2}
\end{equation}
for all $\pth>0$ and for all $\spe$.
\end{proof}

Let us now discuss the meaning and implications
of Eqs.~\eqref{eq:pseudo_eq1} and~\eqref{eq:pseudo_eq2}.

First,
{the mechanism 
underpinning the dynamics of pseudo detailed balanced CRNs
as well as the emergence of self-organized dissipative structures 
is the minimization of $\freepdb$
which plays the role of a Lyapunov function.}
However, $\freepdb$ is not a thermodynamic potential,
but rather a kinetic potential: 
the chemical potentials~\eqref{eq:scp} encode kinetic properties,
namely, the symmetric part of the reaction fluxes $\{\sr{}\}$ 
via $\{\shx\}$ (see Eqs.~\eqref{eq:delta_def} and~\eqref{eq:bbp}).
 
Second,
Eqs.~\eqref{eq:pseudo_eq1} and~\eqref{eq:pseudo_eq2} 
physically mean that diffusion processes are equilibrated at steady state.
This, together with Eq.~\eqref{eq:teqdff} and the fact that $\cpxe $ and $\cpx $ are related through the constant coefficient $\shx$, see Eq.~\eqref{eq:scp},
implies
\begin{equation}
\ds\cpss  = 0\quad\forall \spe\,,
\label{eq:dcph_pdb}
\end{equation}
namely, the chemical potentials of all species are homogeneous.

Third, spatial self-organization at steady state leads, in general, to inhomogeneous reaction fluxes.
Indeed, as long as reaction fluxes do not exclusively depend on the chemical potentials,
homogeneous chemical potentials do not imply homogeneous reaction fluxes.
However, this is not the case for, for example, Arrhenius-like reaction fluxes~\eqref{eq:arr_gen} which become homogeneous when the chemical potentials are homogeneous (see also Subs.~\ref{subs:ex_pdb_m}).

Fourth,
by plugging $\dcurrss = 0$ into Eq.~\eqref{eq:rdeq}, we obtain that
the concentration fields $\convss$ are also a steady state of the chemical dynamics for the internal species in every point $\pos\in V$, i.e., 
\begin{equation}
\sum_{\rct>0}\matSx{}\,\rcurrss{} =0\,.
\end{equation}
Since both diffusion and the chemical reactions are at steady state, 
no oscillations or moving structures (e.g., chemical waves) can emerge.

Fifth,
$\convss$ is not, in general, an equilibrium steady state: 
diffusion processes are equilibrated, i.e., $\eprdffss = 0$,
but chemical reactions are not.
Indeed, $\eprrctess$ is not the thermodynamic reaction entropy production rate~\eqref{eq:eprrct1}
and $\eprrctess = 0$ in Eq.~\eqref{eq:pseudo_eq1} does not imply $\eprrctss = 0$.
The expressions of $\eprrctess$ in Eq.~\eqref{eq:eprrcte2} and $\eprrctss$ in Eq.~\eqref{eq:eprrct2} have a similar structure, given in terms of the lumped reaction fluxes $\hat\omega_\varepsilon$ and the actual reaction fluxes $\omega_\rho$, respectively. Yet, $\eprrctess$ does not resolve the dissipation of the single reactions $\rct\in\setrctpth{}$ in the \pathway{} $\pth$, and it is always a lower bound of the $\eprrctss$ (because of the log sum inequality).

Sixth, by re-writing the \pathway{} net current $\pcurr{}$ according to Eq.~\eqref{eq:pcurr3} 
with fluxes satisfying the pseudo local detailed balance condition~\eqref{eq:ldb_pb}, 
the RD dynamics~\eqref{eq:rdeq} for the internal species is mathematically equivalent to 
the RD dynamics of a detailed balanced (and hence driven by purely passive mechanisms) RD system.
For this reason, we named this kind of CRNs pseudo detailed balanced.
Furthermore, 
Eq.~\eqref{eq:pseudo_eq2} implies that $(\cpgss{\spex} - \shx)$ can be written as 
a linear combination of unbroken conservation laws similar to what happens at equilibrium 
where the equilibrium chemical potentials can be written as a linear combination of conservation laws 
(see Eq.~\eqref{eq:teqrct} and App.~\ref{app:free_var}).

We conclude this discussion by showing that the \pathway{} net currents $\pcurr{}$ 
cannot, in general, be written as in Eq.~\eqref{eq:pcurr3} 
when CRNs are not pseudo detailed balanced.
Namely, there are no $\{\shx\}$ satisfying  Eq.~\eqref{eq:delta_def}.
Indeed, if there were,
the existence of a right-null vector of $\matSxe{}$ as defined in Eq.~\eqref{eq:path_balanced} would imply
\begin{equation}
\sum_{\pth>0}\ln\bbp{}\cyclee = 0\,.
\label{eq:condition_no_mapping}
\end{equation}
However, Eq.~\eqref{eq:condition_no_mapping} cannot hold in general since 
$\{\bbp{}\}$ is defined using the symmetric part of the reaction fluxes so encoding kinetic properties, 
while $\{\cyclee\}$ is derived from the stoichiometric matrix so encoding topological properties only.

In summary, the nonequilibrium dynamics of pseudo detailed balanced CRNs, as well as the resulting self-organized structures,
can be exactly obtained by passive mechanisms too.

\subsection{Complex Balanced CRNs\label{sub:cb}}
CRNs are said to be complex balanced~\cite{Anderson2010} when there exists an homogeneous steady state $\convssh$ of the chemical dynamics, 
i.e., $\sum_{\rct>0}\matSx{}\rcurrssh{}=0$, 
such that 
the sum of the net currents entering into a complex $\com$ equals the  sum of the net currents exiting from it,
namely, 
$\sum_{\rct>0} \matIe \rcurrssh{}=0$ $\forall\com$,
or equivalently 
\begin{equation}
\sum_{\rct} \dk_{\com,\com(\rct)} \rcurrssh{}=0 \text{ }\text{ }\text{ }\forall\com\,.
\label{eq:cb}
\end{equation}
Deficiency-zero CRNs, 
where every cycle (i.e., every right-null eigenvector of the substoichiometric matrix for the internal species)
is also a right-null eigenvector of the incidence matrix, are always complex balanced.
If deficiency is greater than zero, then CRNs are, in general, not complex balanced 
(except for a set of kinetic parameters of null measure~\cite{Cappelletti2016,Polettini2015}).

We now show that the RD dynamics of complex balanced CRNs 
is driven by the minimization of a kinetic potential and
relaxes to a steady state $\convss$ characterized by a vanishing diffusion entropy production rate, i.e., $\eprdffss = 0$, 
independently of whether the concentration fields are homogeneous or inhomogeneous.
We do so 
by generalizing a previous proof for the existence of a Lyapunov function for homogeneous complex balanced CRNs in dilute solutions undergoing mass-action kinetics~\cite{Anderson2014_Lyapunov}.

\assumption
We consider reaction fluxes of Arrhenius-like form~\eqref{eq:arr_gen} and we rewrite them as
\begin{equation}
\rflux{} = \Ap{}\,e^{\frac{\sum_{\spex}\cpx\stocx{}}{RT}}\,,
\label{eq:arr}
\end{equation} 
with $\Ap{} \equiv \AAp{}\exp({({\sum_{\spey}\cpyy\stocy{})}/{RT}}) \neq \Ap{-}$ being a constant parameter accounting (also) for the chemostatted species. 
Note that, 
since $\stocxc{(\rct)}=\stocx{}$ by definition of complexes, 
Eq.~\eqref{eq:cb} becomes
\begin{equation}
\sum_{\rct}\dk_{\com,\com(\rct)}
\Big\{
\Ap{} -\Ap{-}e^{\frac{\sum_{\spex}\cpxssh\matSx{}}{RT}}
\Big\}=0\,,
\label{eq:cb2}
\end{equation}
when endowed with Arrhenius-like reaction fluxes~\eqref{eq:arr}.

\begin{proof}
We start by introducing the potential  
\begin{equation}
\freecb \equiv \free - \sum_{\spex}\cpxssh\int_V\dd\pos\,\conx\,,
\label{eq:LYcb}
\end{equation}
and then we show that $\freecb$ acts as a Lyapunov function of the RD dynamics~\eqref{eq:rdeq} with respect to 
a steady state characterized by homogeneous chemical potentials.
Note that the potential~\eqref{eq:LYcb} is obtained from the free energy~\eqref{eq:free} 
via a shift of the standard chemical potentials of the internal species $\{\stcpx\}$ given by
\begin{equation}
\shx = - \cpxssh \,,
\label{eq:shift:cb}
\end{equation}
like what is done in Eq.~\eqref{eq:scp} for pseudo detailed balanced CRNs. 

By using Eqs.~\eqref{eq:rdeq},~\eqref{eq:arr}, and $\dt\conyy = 0$, the time derivative of $\freecb$ reads
\begin{equation}
\dt \freecb= -T{\eprrctc} - T\eprdff  \,,
\end{equation}
where 
$\eprdff\geq 0$ is given in Eq.~\eqref{eq:eprdff2} (since $\ds \cpyy = 0$) , and
\begin{equation}
\begin{split}
T\eprrctc\equiv -
\int_V\dd\pos \sum_{\rct,\, \spex}\Big\{
(&\cpg{\spex} - \cpgssh{\spex}) \\
&\matSx{} \Ap{} e^{\frac{\sum_{\spex}\cpx\stocx{}}{RT}}\Big\}\,.
\end{split}
\end{equation}

We have now to show that $-\eprrctc\leq 0$. 
By using 
i) Eq.~\eqref{eq:matS},
ii) multiplying and dividing by $\exp{(\sum_{\spex}\cpxssh\stocx{}/{RT})}$,
and iii) $e^\rna(\rnb - \rna) \leq e^\rnb - e^\rna$ for any real number $\rna$ and $\rnb$
(where the equality sign is achieved if and only if $\rna = \rnb$), we obtain
\begin{equation}\small
\begin{split}
-\frac{\eprrctc}{R}&\leq
\int_V\dd\pos \sum_{\rct}
\Big\{
e^{\frac{\sum_{\spex}(\cpx - \cpxssh)\stocx{-}}{RT}} 
\\
-
&e^{\frac{\sum_{\spex}(\cpx - \cpxssh)\stocx{}}{RT}}
\Big\}
\Ap{} e^{\frac{\sum_{\spex}\cpxssh\stocx{}}{RT}}\,,
\end{split}
\label{eq:cb_inter}
\end{equation}
where the equality sign is achieved if and only if $(\cpg{\spex} - \cpgssh{\spex}) \matSx{}=0 $.
Since $\sum_{\rct} = \sum_{\com}\sum_{\rct}\dk_{\com,\com(\rct)}$ and $\stocxc{(\rct)}=\stocx{}$ by definition of complexes, 
the two integrands in Eq.~\eqref{eq:cb_inter} can be rewritten as
\begin{subequations}
\begin{align}
&\sum_{\com}e^{\frac{\sum_{\spex}\cpx\stocxc{}}{RT}}\sum_{\rct}\dk_{\com,\com(\rct)}\Ap{-}e^{\frac{\sum_{\spex}\cpxssh\matSx{}}{RT}}\,,\\
&\sum_{\com}e^{\frac{\sum_{\spex}\cpx\stocxc{}}{RT}}\sum_{\rct}\dk_{\com,\com(\rct)}\Ap{}\,,
\end{align}
\end{subequations}
respectively,
and hence the r.h.s. of Eq.~\eqref{eq:cb_inter} reads
\begin{equation}
\begin{split}
-\int_V\dd\pos \sum_{\com}&e^{\frac{\sum_{\spex}\cpx\stocxc{}}{RT}}
\sum_{\rct}\dk_{\com,\com(\rct)}
\Big\{
\Ap{} 
\\
&
-\Ap{-}e^{\frac{\sum_{\spex}\cpxssh\matSx{}}{RT}}
\Big\}=0
\,,
\end{split}
\end{equation}
because of Eq.~\eqref{eq:cb2}.
Thus, 
\begin{equation}
\dt \freecb\leq 0\,.
\label{eq:lyapunov_cb}
\end{equation}
This, together with the fact that 
$\freecb$ is lower bounded (since $\free$ is lower bounded by assumption, see Subs.~\ref{subs:thermo_cost}),
implies that the RD dynamics~\eqref{eq:rdeq} of complex balance CRNs must relax towards a steady state $\convss$ satisfying
\begin{equation}
\eprrctcss = 0 \text{ }\text{ }\text{ }\text{ and }\text{ }\text{ }\text{ } \eprdffss = 0\,,
\label{eq:complex_eq1}
\end{equation}
or equivalently
\begin{equation}\small
\sum_{\spex}
(\cpgss{\spex} - \cpgssh{\spex}) \matSx{}=0 \text{ }\text{ }\text{ }\text{ and }\text{ }\text{ }\text{ }\dcurrss = 0 \,,
\label{eq:complex_eq2}
\end{equation}
for all $\rct>0$ and for all $\spe$.
\end{proof}

As we did in Subs.~\ref{sub:pdb}, we now further discuss the conditions, the physical meaning, and the implications
of Eqs.~\eqref{eq:complex_eq1} and~\eqref{eq:complex_eq2}.

First,
as for pseudo detailed balanced CRNs,
{the mechanism 
underpinning the dynamics of complex balanced CRNs
as well as the emergence of self-organized dissipative structures 
is the minimization of $\freecb$ in Eq.~\eqref{eq:LYcb}
which plays the role of a Lyapunov function.}
However, $\freecb$ is not a thermodynamic potential,
but rather a kinetic potential: 
it encodes kinetic properties via the steady state concentrations in $\{\cpxssh\}$ (see Eq.~\eqref{eq:LYcb}).

Second,
as for pseudo detailed balanced CRNs,
Eqs.~\eqref{eq:complex_eq1} and~\eqref{eq:complex_eq2} 
physically mean that diffusion processes are equilibrated at steady state.
This, together with Eq.~\eqref{eq:eprdff2},
implies that the chemical potentials of all species are homogeneous.

Third,
spatial self-organization at steady state does not lead to inhomogeneous reaction fluxes.
Indeed, homogeneous chemical potentials at steady state, 
together with reaction fluxes of Arrhenius-like form~\eqref{eq:arr}, imply that
the reaction fluxes are homogeneous too, unlike for pseudo detailed balanced CRNs.

Fourth,
as for pseudo detailed balanced CRNs, by plugging $\dcurrss = 0$ into Eq.~\eqref{eq:rdeq},  
we obtain that both diffusion processes and chemical reactions are at steady state 
in every point $\pos\in V$,.
Consequently, no oscillations or moving structures (e.g., chemical waves) can emerge.

Fifth,
as for pseudo detailed balanced CRNs, 
$\convss$ is not, in general, an equilibrium steady state
even if the diffusion processes are equilibrated, i.e., $\eprdffss = 0$.

Sixth,
Eq.~\eqref{eq:complex_eq2} implies that $(\cpgss{\spex} - \cpgssh{\spex})$ can be written as 
a linear combination of unbroken conservation laws similarly to what happens at equilibrium 
where the equilibrium chemical potentials can be written as a linear combination of conservation laws 
(see Eq.~\eqref{eq:teqrct} and App.~\ref{app:free_var}).

In summary, the nonequilibrium dynamics of complex balanced CRNs, as well as the resulting self-organized structures,
shares important features with passive mechanisms, 
e.g., it is driven by the minimization of (appropriately constructed) potentials and diffusion processes equilibrate.

\subsection{Generalization to Non-ideal Chemostatted Species\label{sub:pdb_cp_nonideal}}
We show here that the results of Subs.~\ref{sub:pdb} and~\ref{sub:cb} can be generalized to the case with non-ideal chemostatted species.
Namely, when the chemical potentials $\{\cpy\}$ 
(given in Eq.~\eqref{eq:cp} and not in Eq.~\eqref{eq:cp_id})
are still homogeneous and constant in time,
i.e., \begin{equation}
\cpy = \cpyy 
\quad\forall\pos \text{ and }\forall t\,,
\label{eq:cpy_constraint}
\end{equation}
but the concentrations $\{\cony\}$ are not.
Note that this implies that the exchange currents $\{\ecurry\}$ in Eq.~\eqref{eq:rdeq} 
represent the mechanisms ensuring that Eq.~\eqref{eq:cpy_constraint} is satisfied, 
but $\pt \cony \neq 0$ in general.

To do so, we recognize that the exchange currents $\{\ecurry\}$
can always be represented as the (net) result of exchange reactions $\rcty\in\setrcte=\{\pm1, \pm2,\dots,\pm\nrcte\}$
between the species $\spey\in\setspey$ and ideal chemostatted species $\speyi\in\setspeyi$ according to
\begin{equation}
\chemy \ch{ <=>[$\rcty$][$-\rcty$]} \chemyi\,.
\label{eq:exrct}
\end{equation}
Indeed, as long as the exchange reactions $\rcty\in\setrcte$ are fast enough 
(compared to the time scale of the other reactions $\rct\in\setrct$ and of the diffusion processes),
they are always equilibrated and, consequently,  Eq.~\eqref{eq:cpy_constraint} is satisfied 
(with $\cpyy$ playing the role of the chemical potential of $\speyi$).
This implies that RD systems with non-ideal chemostatted species are equivalent to 
RD systems with $\setspex\bigcup\setspey\bigcup\setspeyi$ species
(where $\setspex\bigcup\setspey$ are internal species and $\setspeyi$ are ideal chemostatted species)
and $\setrct\bigcup\setrcte$ reactions.
As proven in Subs.~\ref{sub:pdb} and~\ref{sub:cb}, 
the latter will relax to a nonequilibrium steady state with homogeneous chemical potentials 
(i.e., where diffusion processes have equilibrated)
if their CRNs are either pseudo detailed balanced or complex balanced
and, therefore, also the former will relax to the same steady state.
Note that the CRN topology of RD systems with $\setspex\bigcup\setspey\bigcup\setspeyi$ species 
might be different from the one of the original RD systems because of the introduction of the exchange reactions~\eqref{eq:exrct}.

\subsection{Emergence of Spatial Structures by Diffusive Processes\label{subs:diff_induced_emergence_self_org}}

In pseudo detailed balanced and complex balanced CRNs,
the steady state spatial structure $\convss$ is determined by the minimization of appropriately constructed kinetic potentials 
obtained by a simple shift of the standard chemical potentials of the internal species $\{\stcpx\}$. 
This, together with the equilibration of the diffusion processes at steady state, 
{implies} that the chemical reactions {are the mechanism determining} the total abundances of the chemical species, 
i.e., $\int_V\dd\pos\,\convss$, 
while diffusion {is the mechanism determining} the spatial structures, i.e., the $\pos$-dependence of $\convss$.
Indeed, as proved in App.~\ref{app:dff_var}, 
the steady-state concentration {profiles} $\convss$ of pseudo detailed balanced and complex balanced CRNs can also be obtained by minimizing the free energy of pure diffusion, $\free$ in Eq.~\eqref{eq:free}, 
when the total abundances are set to values equal to $\int_V\dd\pos\,\convss$.

{In CRNs that are neither pseudo detailed balanced nor complex balanced,
this differentiation between the roles of chemical reactions and diffusion no longer holds in general
and both mechanisms contribute to determining the spatial structures.}


\subsection{Final Comment\label{sub:final_comments}}
We notice that the equilibration of diffusion processes in complex balanced CRNs endowed with Arrhenius-like fluxes
was also recently derived in Ref.~\cite{Miangolarra2023} with a different approach.
Let us now stress the differences between our work and Ref.~\cite{Miangolarra2023}.
First,
our derivation is a closed derivation at the deterministic level of description of the RD dynamics.
Their derivation is based on the stochastic dynamics of complex balanced CRNs.
Second,
our derivation is consistent with nonlocal terms 
(e.g., of the form $\ds \conx \cdot \matKij \ds \conxx$, see Eq.~\eqref{eq:ex:free} in Sec.~\ref{sec:num})
in the non-ideal free energy.
Their derivation determines the Lyapunov function of the reaction dynamics (the equivalent of $\freecb$ in Eq.~\eqref{eq:LYcb})
by taking the macroscopic limit of the steady-state rate function for \textit{homogeneous} CRNs.
Hence, their Lyapunov function can consistently include only local terms 
(e.g., of the form $\conx \matMij \conxx$, see Eq.~\eqref{eq:ex:free} in Sec.~\ref{sec:num})
 in the non-ideal free energy 
(although they use nonlocal terms in one of the examples).
Third,
our derivation can be applied to RD systems with non-ideal chemostatted species (see Subs.~\ref{sub:pdb_cp_nonideal}).
Their derivation requires ideal chemostatted species.

%

\section{Thermodynamic Properties of Active Self-Organization - Numerical Results\label{sec:num}}
We now examine the thermodynamic properties (introduced in Sec.~\ref{sec:thermo}) 
of specific RD systems.
We first illustrate the exact results obtained in Sec~\ref{sec:aps} 
for pseudo detailed balanced CRNs
and complex balanced CRNs.
We then examine a CRN that is neither pseudo detailed balanced nor complex balanced.
To do so, we will numerically simulate their RD dynamics~\eqref{eq:rdeq}
to determine the final steady state $\convss$ 
starting from a homogeneous steady state $\convssh$ 
after the concentration of each internal species $\spex$ is perturbed
with a gaussian white noise of zero average and $(0.1*\conxh)^2$~variance.
We consider RD systems in a two-dimensional space with periodic boundary conditions.
Simulations are performed using the py-pde package developed in Ref.~\cite{py-pde}.

\subsection{Model}
We consider RD systems where
the set of internal species includes a nonreacting species, labeled $\nr$ in the following,
while the chemostatted species are assumed to be ideal, 
with homogeneous and time-independent chemical potentials.

The non-ideal free energy,
accounting for the interactions between the internal species only,
is assumed to have the form 
\begin{equation}
\begin{split}
\freeni = \frac{RT}{2}\sum_{\spex,\spexx} \int_V\dd\pos\, 
\big[
&\conx \matMij \conxx 
\\
&+\ds \conx \cdot \matKij \ds \conxx
\big]\,,
\end{split}
\label{eq:ex:free}
\end{equation}
where $\{\matMij\}$ and $\{\matKij\}$ are the entries of the two (symmetric) matrixes $\matM$ and $\matK$ describing 
the mean-field molecular interactions and the cost of forming interfaces, respectively.
{If $\matMij > 0$ (resp. $\matMij < 0$ ), then $\matM$ represents repulsive (resp. attractive) interactions between 
$\spex$ and $\spexx$:
large concentrations $\conx$ and $\conx$ in the same region increase (resp. decrease) the free energy,
i.e., $\conx \matMij \conxx > 0$ (resp. $\conx \matMij \conxx < 0$).
Similarly,
if $\matKij > 0$ (resp. $\matKij < 0$ ), then $\matK$ represents an energetic advantage (resp. disadvantage) 
in distributing $\spex$ and $\spexx$ in different regions:
gradients $\ds\conx$ and $\ds\conx$ with opposite sign at the interface 
between a region enriched in $\spex$ and a region enriched in $\spexx$ decrease (resp. increase) the free energy,
i.e., $\ds \conx \cdot \matKij \ds \conxx < 0$ (resp. $\ds \conx \cdot \matKij \ds \conxx < 0$).
Note that the non-ideal free energy in Eq.~\eqref{eq:ex:free} 
is consistent with the one used in Refs.~\cite{Saha2020,Joanny2020} 
and corresponds to one introduced by Cahn and Hilliard~\cite{cahn1958free} 
if higher orders in gradients are included.
The corresponding chemical potentials of the internal species are given by
}
\begin{equation}
\begin{split}
\cpg{\spex} = \stcpx &+ RT \ln \conx 
\\
&+RT \sum_{\spexx}\big[\matMij \conxx  -  \matKij \nabla^2 \conxx\big] \,,
\end{split}
\end{equation}
where we also used Eqs.~\eqref{eq:cp} and~\eqref{eq:f_ideal}.
The chemical potentials of the chemostatted species $\{\cpyy\}$ are treated as control parameter like in Sec.~\ref{sec:aps}
and are related to the corresponding concentrations $\{\conyy\}$ via Eq.~\eqref{eq:cp_id}.

Reaction fluxes are assumed to follow Arrhenius-like forms given in Eq.~\eqref{eq:arr}.
The entries of the Onsager matrix $\matO$ in the diffusion currents~\eqref{eq:dcurr} are assumed to read
\begin{equation}
\matOij = \Di \conc \dk_{\spe,\spee}\,,
\end{equation}
where $\Di$ is the diffusion coefficient of species $\spe$.
Hence, the diffusion currents for the internal species are given by
\begin{equation}
\dcurrx = -\Dx \conx \ds \cpg{\spex} \text{ } \text{ }\forall\spex\in\setspex\,.
\end{equation}

\subsection{Pseudo Detailed Balanced CRN \label{subs:ex_pdb_m}}
We consider a RD system with 
three internal species $\{\ch{X_1}, \ch{X_2}, \ch{X_{\nr}}\}$ 
and two chemostatted species $\{\ch{Y_1}, \ch{Y_2}\}$
which are interconverted via the reactions
\begin{equation}
\begin{split}
X_1+Y_1 &\ch{ <=>[  $1$ ][ $-1$ ]} X_2 + Y_2\,,\\
X_1 &\ch{ <=>[  $2$ ][  $-2$ ]} X_2\,.
\end{split}\label{eq:pdb-example}
\end{equation}
The CRN~\eqref{eq:pdb-example} has three complexes $V_1 = \ch{X_1}$, $V_2=\ch{X_2}$, and $V_3 = \ch{X_\nr}$
and the corresponding graph of complexes reads
\begin{equation}
\begin{split}
V_1 &\ch{ <=>[  $1$ ][ $-1$ ]} V_2\,,\\
V_1 &\ch{ <=>[  $2$ ][  $-2$ ]} V_2\,,
\end{split}
\end{equation}
admitting only the \pathway{} from $V_1$ to $V_2$ and its backward counterpart from $V_2$ to $V_1$.
Hence, the stoichiometric matrix for the internal species,
\begin{equation}
{\matSX}=
 \kbordermatrix{
    & \color{g}1 &\color{g}2\cr
    \color{g}\ch{X_1} 	   &-1  &-1  \cr
    \color{g}\ch{X_2}  	   & 1  & 1  \cr
    \color{g}\ch{X_\nr}   & 0  & 0 \cr
  }\,,
  \label{eq:example1_matS}
\end{equation}
becomes the matrix 
\begin{equation}
  \kbordermatrix{
    & \color{g} \cr
    \color{g}\ch{X_1} 	   &-1   \cr
    \color{g}\ch{X_2}  	   & 1   \cr
    \color{g}\ch{X_\nr}   & 0    \cr
  }\,,
  \label{eq:example1_matS_ptw}
\end{equation}
when considering the only \pathway.
Since the matrix in Eq.~\eqref{eq:example1_matS_ptw} does not admit any right-null eigenvector,
the CRN~\eqref{eq:pdb-example} is pseudo detailed balanced (see Subs.~\ref{sub:pdb}).

\begin{figure}
    \centering
    \includegraphics[width=0.49\textwidth]{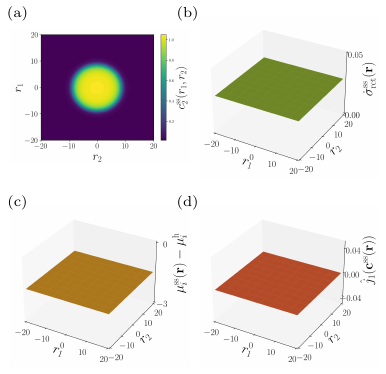}
    \caption{Steady state and corresponding thermodynamic properties
     of the pseudo detailed balanced CRN~\eqref{eq:pdb-example}.  
    (a)~Concentration field of~$\ch{X_2}$. 
    (b)~Density of reaction entropy production rate.
    (c)~Difference between the chemical potential of $\spe\in\{\ch{X_1},\ch{X_2}\}$ (labeled here $\mu_i^{\text{ss}}(\pos)$)
    and its chemical potential at the homogeneous steady state (labeled here $\mu_i^{\text{h}}$). 
    (d)~Net current of the \pathway, i.e., $\hat{j}_1(\convss) = j_1(\convss) + j_2(\convss)$. 
    Simulation parameters in arbitrary units: 
    $RT = 1$,
    $\chi=3$, $k_1=0.1$, $k_2=0.05$, 
    $\mu_1^\circ=0$, $\mu_2^\circ=-2$, $\mu_{\ch{Y}_1}=1$, $\mu_{\ch{Y}_2}=0$, 
    $A_\rct = 1$ $\forall \rct$, 
    $D_x=1$ $\forall x$, 
    $c_\text{1}^{\mathrm{h}}= 0.087$, $c_\text{2}^{\mathrm{h}}= 0.213$, $c_\text{nr}^{\mathrm{h}}=0.7$. 
    }
    \label{fig:pdb}
\end{figure}

We solve the RD dynamics~\eqref{eq:rdeq}
when the mean-field molecular interactions and the cost of forming interfaces are given by the following matrixes
\begin{align}
\label{eq:interaction-matrices-pdb}
&\mathbb{M} =
\kbordermatrix{
& \color{g} \ch{X_1} & \color{g} \ch{X_2}   &\color{g} \ch{X_\nr}\cr
\color{g} \ch{X_1}         & 0 & \chi & 0  \cr
\color{g} \ch{X_2}        & \chi & 0 & \chi \cr
\color{g} \ch{X_\nr}     & 0 & \chi & 0 \cr
}
\,,\quad
&\mathbb{K} =
\kbordermatrix{
& \color{g} \ch{X_1} & \color{g} \ch{X_2}   &\color{g} \ch{X_\nr}\cr
\color{g} \ch{X_1} & k_1 & k_2 & 0  \cr
\color{g} \ch{X_2} & k_2 & k_1 & k_2  \cr
\color{g} \ch{X_\nr} & 0 & k_2 & 0 \cr
}\,.
\end{align}
The typical steady-state solution and its thermodynamic properties are shown in Fig.~\ref{fig:pdb}.
We observe that the concentration fields spatially organize (Fig.~\ref{fig:pdb}a) into a pattern 
that is qualitatively analogous to a complete phase separation obtained at equilibrium:
the species \ch{X_2} accumulates in a \textit{single droplet}.
The chemical reactions are however out of equilibrium, i.e., $\deprrctss>0$ (see Fig.~\ref{fig:pdb}b),
even if the net current along the \pathway, namely, 
$\hat{j}_1(\convss) = j_1(\convss) + j_2(\convss)$, vanishes as predicted in Subs.~\ref{sub:pdb} (see Fig.~\ref{fig:pdb}d).
Furthermore, in agreement with our derivation in Subs.~\ref{sub:pdb}, at steady state
diffusion processes are at equilibrium, i.e., $\deprdffss=0$,
because the chemical potentials are homogeneous (see Fig.~\ref{fig:pdb}c).

Note that the steady-state density of reaction entropy production rate in Fig.~\ref{fig:pdb}b is homogeneous  
because we used reaction fluxes of Arrhenius-like forms, 
and, therefore the reactions fluxes become homogeneous at steady state (as explained in Subs.~\ref{sub:pdb}).

\subsection{Complex Balanced CRN\label{subs:example_cb_solution}}
We consider a RD system with 
four internal species $\{\ch{X_1}, \ch{X_2}, \ch{X_3}, \ch{X_{\nr}}\}$ 
and two chemostatted species $\{\ch{Y_1}, \ch{Y_2}\}$
which are interconverted via the reactions
\begin{equation}
\begin{split}
        X_1+Y_1 &\ch{ <=>[  $1$ ][ $-1$ ]} X_2 + Y_2\,,\\
        X_2 &\ch{ <=>[  $2$ ][  $-2$ ]} X_3\,,\\
        X_3 &\ch{ <=>[  $3$ ][  $-3$ ]} X_1\,.
        \end{split}
        \label{eq:cb-reactions}
\end{equation}
The CRN~\eqref{eq:cb-reactions} has four complexes $V_1 = \ch{X_1}$, $V_2=\ch{X_2}$, $V_3=\ch{X_3}$, and $V_4=\ch{X_\nr}$
and the corresponding graph of complexes reads
\begin{equation}
\begin{split}
V_1 &\ch{ <=>[  $1$ ][ $-1$ ]} V_2\,,\\
V_2 &\ch{ <=>[  $2$ ][  $-2$ ]} V_3\,,\\
V_3 &\ch{ <=>[  $3$ ][  $-3$ ]} V_1\,.
\end{split}
\end{equation}
Since the substoichiometric matrix for the internal species
\begin{equation}
{\matSX}=
 \kbordermatrix{
    & \color{g}1 &\color{g}2&\color{g}3\cr
    \color{g}\ch{X_1} 	   &-1  & 0  & 1  \cr
    \color{g}\ch{X_2}  	   & 1  &-1  & 0 \cr
    \color{g}\ch{X_3}  	   & 0  & 1  &-1 \cr
    \color{g}\ch{X_\nr}   & 0  & 0  & 0\cr
  }\,
  \label{eq:example2_matS}
\end{equation}
admits only one cycle that is also a right-null eigenvector of the incidence matrix, 
\begin{equation}
{\matI}=
 \kbordermatrix{
    & \color{g}1 &\color{g}2&\color{g}3\cr
    \color{g}\ch{V_1} 	   &-1  & 0  & 1  \cr
    \color{g}\ch{V_2}  	   & 1  &-1  & 0 \cr
    \color{g}\ch{V_3}  	   & 0  & 1  &-1 \cr
    \color{g}\ch{V_4}  	   & 0  & 0  & 0 \cr
  } \,,
  \label{eq:example2_matI}
\end{equation}
the CRN~\eqref{eq:cb-reactions} is deficiency-zero (see Eq.~\eqref{eq:dim}) and, therefore, complex balanced (see Subs.~\ref{sub:cb}).

\begin{figure}
    \centering\includegraphics[width=0.49\textwidth]{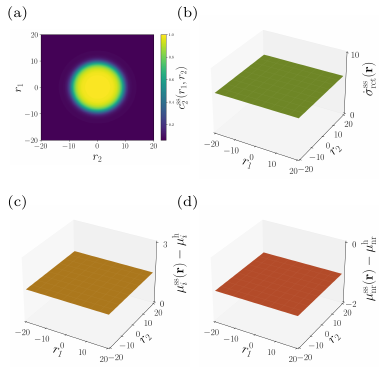}
    \caption{Steady state and corresponding thermodynamic properties
     of the complex balanced CRN~\eqref{eq:cb-reactions}. 
    (a)~Concentration field of $\ch{X_2}$. 
    (b)~Density of reaction entropy production rate.
    (c)~Difference between the chemical potential of $\spe\in\{\ch{X_1},\ch{X_2},\ch{X_3}\}$ 
     (labeled here $\mu_i^{\text{ss}}(\pos)$)
    and its chemical potential at the homogeneous steady state (labeled here $\mu_i^{\text{h}}$). 
    (d)~Difference between the chemical potential of \ch{X_\nr} (labeled here $\mu_\nr^{\text{ss}}(\pos)$)
    and its chemical potential at the homogeneous steady state (labeled here $\mu_\nr^{\text{h}}$).
    Simulation parameters in arbitrary units: 
     $RT = 1$,
    $\chi=3$, $k_1=0.5$, $k_2=0.1$, 
    $\mu_{1,2}^{\circ}=1$, $\mu_{3}^{\circ}=3$, $\mu_{\text{nr}}^{\circ}=0$, $\mu_{Y_1}=4$, $\mu_{Y_2}=0$, 
    $A_\rct = 1$ $\forall \rct$, 
    $D_x=1$ $\forall x$, 
    $c_\text{1}^{\mathrm{h}}= 0.007$, $c_\text{2}^{\mathrm{h}}= 0.274$, $c_\text{3}^{\mathrm{h}}= 0.019$, $c_\text{nr}^{\mathrm{h}}=0.7$. 
    }
    \label{fig:cb-solution}
\end{figure}

We solve the RD dynamics~\eqref{eq:rdeq}
when the mean-field molecular interactions and the cost of forming interfaces are given by the following matrixes
\begin{align}
\label{eq:interaction-matrices-cb}
&\mathbb{M}=
\kbordermatrix{
  & \color{g} \ch{X_1} & \color{g} \ch{X_2} & \color{g}  \ch{X_3}  &\color{g} \ch{X_\nr}\cr
\color{g} \ch{X_1} & 0 & \chi & 0 & 0 \cr
\color{g} \ch{X_2} & \chi & 0 & \chi & \chi \cr
\color{g} \ch{X_3} & 0 & \chi & 0 & 0 \cr
\color{g} \ch{X_\nr} & 0 & \chi & 0 & 0 \cr
}
\,,
&\mathbb{K} =
\kbordermatrix{
  & \color{g} \ch{X_1} & \color{g} \ch{X_2} & \color{g}  \ch{X_3}  &\color{g} \ch{X_\nr}\cr
\color{g} \ch{X_1} & k_1 & k_2 & 0 & 0 \cr
\color{g} \ch{X_2} & k_2 & k_1 & k_2 & k_2 \cr
\color{g} \ch{X_3} & 0 & k_2 & k_1 & 0  \cr
\color{g} \ch{X_\nr} & 0 & k_2 & 0 & k_1 \cr
}\,.
\end{align}
The typical steady-state solution and its thermodynamic properties are shown in Fig.~\ref{fig:cb-solution}.
As for the RD system in Subs.~\ref{subs:ex_pdb_m},
we observe that the concentration fields spatially organize (Fig.~\ref{fig:cb-solution}a) into a pattern 
that is qualitatively analogous to a complete phase separation obtained at equilibrium:
the species \ch{X_2} accumulates in a \textit{single droplet}.
The chemical reactions are however out of equilibrium, i.e., $\deprrctss>0$ (see Fig.~\ref{fig:cb-solution}b),
while diffusion processes are at equilibrium, i.e., $\deprdffss=0$,
since the chemical potentials are homogeneous (see Fig.~\ref{fig:cb-solution}c) as predicted by our derivation in Subs.~\ref{sub:cb}.

Furthermore, according to Eq.~\eqref{eq:complex_eq2}, 
the difference $(\cpgss{\spex} - \cpgssh{\spex})$ for every internal species 
can be written as a linear combination of unbroken conservation laws (defined in Subs.~\ref{sub:claw}).
This implies that in general
\begin{equation}
\begin{split}
(\cpgss{\ch{X_1}} - \cpgssh{\ch{X_1}})
&= (\cpgss{\ch{X_2}} - \cpgssh{\ch{X_2}})\\
&= (\cpgss{\ch{X_3}} - \cpgssh{\ch{X_3}})\\
&\neq(\cpgss{\ch{X_\nr}} - \cpgssh{\ch{X_\nr}})\,,
\end{split}
\label{eq:cb1_constr_cp}
\end{equation}
since the stoichiometric matrix for the internal species~\eqref{eq:example2_matS} admits the following left null eigenvectors
\begin{equation}
 \kbordermatrix{
    & \color{g}\cr
    \color{g}\ch{X_1} 	   & 1 \cr
    \color{g}\ch{X_2}  	   & 1 \cr
    \color{g}\ch{X_3}  	   & 1 \cr
    \color{g}\ch{X_\nr}   & 0 \cr
  }\,,
  \quad\quad\quad
\kbordermatrix{
    & \color{g}\cr
    \color{g}\ch{X_1} 	   & 0 \cr
    \color{g}\ch{X_2}  	   & 0 \cr
    \color{g}\ch{X_3}  	   & 0 \cr
    \color{g}\ch{X_\nr}   & 1 \cr
  }\,.
\end{equation}
The constraint in Eq.~\eqref{eq:cb1_constr_cp} is confirmed by the simulation
(compare Fig.~\ref{fig:cb-solution}(c) and Fig.~\ref{fig:cb-solution}(d)).

\subsection{Brusselator\label{subs:brusselator}}
We consider a RD system of 
three internal species $\{\ch{X_1}, \ch{X_2}, \ch{X_{\nr}}\}$ 
and four chemostatted species $\{\ch{Y_1}, \ch{Y_2}, \ch{Y_3}, \ch{Y_4}\}$
which are interconverted via the reactions
\begin{equation}
\begin{split}
        Y_ 1 &\ch{ <=>[ $1$ ][ $-1$ ]} X_ 1 \,,\\
        2X_ 1+X_ 2 &\ch{ <=>[ $2$ ][ $-2$ ]} 3X_ 1\,,\\
        Y_ 2+X_ 1 &\ch{ <=>[ $3$ ][ $-3$ ]} X_ 2+Y_ 4\,,\\
        X_ 1 &\ch{ <=>[ $4$ ][ $-4$ ]} Y_ 3\,.
        \end{split}
        \label{eq:crn-brusselator} 
\end{equation}
The CRN~\eqref{eq:crn-brusselator} is known as Brusselator and 
can be represented in terms of the 6 complexes
$V_1 = \ch{X_1}$, $V_2 = \ch{X_2}$, $V_3 = \ch{2 X_1 + X_2}$, $V_4 = \ch{3 X_1}$, $V_5 = \ch{X_\nr}$, and $\emptyset= Y_{1,2}$
and the following graph of complexes:
\begin{equation}
\begin{split}
        \emptyset &\ch{ <=>[ $1$ ][ $-1$ ]} V_ 1 \,,\\
        V_3 &\ch{ <=>[ $2$ ][ $-2$ ]} V_4\,,\\
        V_ 1 &\ch{ <=>[ $3$ ][ $-3$ ]} V_ 2\,,\\
        V_ 1 &\ch{ <=>[ $4$ ][ $-4$ ]} \emptyset\,.
        \end{split}
        \label{eq:gc-brusselator}
\end{equation}
The stoichiometric matrix for the internal species 
\begin{equation}
{\matSX}=
 \kbordermatrix{
    & \color{g}1 &\color{g}2&\color{g}3&\color{g}4\cr
    \color{g}\ch{X_1} 	   & 1  & 1 & -1  &-1  \cr
    \color{g}\ch{X_2}  	   & 0  &-1  & 1  & 0  \cr
    \color{g}\ch{X_\nr}   & 0  & 0  & 0  & 0  \cr
  }\,
  \label{eq:example_brusselator_matS}
\end{equation}
admits 2 cycles, i.e., 
\begin{equation}
\cyclev=
\kbordermatrix{
   & \cr
   \color{g}\ch{1} 	 	     & 1      \cr
   \color{g}\ch{2}  		     & 0      \cr
   \color{g}\ch{3}  	  	     & 0      \cr
   \color{g}\ch{4}  	  	     & 1      \cr
 }
\quad\text{ and }\quad
\cyclev'=
\kbordermatrix{
   & \cr
   \color{g}\ch{1} 	 	     & 0      \cr
   \color{g}\ch{2}  		     & 1      \cr
   \color{g}\ch{3}  	  	     & 1      \cr
   \color{g}\ch{4}  	  	     & 0      \cr
 }\,,
\end{equation}
but only the first one is a right-null vector of the incidence matrix
\begin{equation}
{\matI}=
 \kbordermatrix{
    & \color{g}1 &\color{g}2&\color{g}3&\color{g}4\cr
    \color{g}\ch{V_1} 	   & 1  & 0  &-1  &-1  \cr
    \color{g}\ch{V_2}  	   & 0  & 0  & 1  & 0  \cr
    \color{g}\ch{V_3}  	   & 0  &-1  & 0  & 0  \cr
    \color{g}\ch{V_4}  	   & 0  & 1  & 0  & 0  \cr
    \color{g}\ch{V_5}  	   & 0  & 0  & 0  & 0  \cr
    \color{g}\emptyset    &-1  & 0  & 0  & 1  \cr
  }\,.
  \label{eq:example_brusselator_matI}
\end{equation}
Hence, the deficiency reads $\deff =1$ (see Eq.~\eqref{eq:dim}) and, in general, the CRN~\eqref{eq:crn-brusselator} is neither pseudo detailed balanced nor complex balanced.
Namely, any homogeneous steady state $\convssh$ of the chemical dynamics satisfying $\sum_{\rct>0}\matSx{}\rcurrssh{}=0$ 
does not satisfy $\sum_{\rct>0} \matIe \rcurrssh{}=0$ too 
(except for a set of parameters, e.g., $\{\AAp{}\}$,  $\{\cpyy\}$, $\matM$, and $\matK$, 
of null measure~\cite{Cappelletti2016,Polettini2015}).

\begin{figure}
    \centering
    \includegraphics[width=0.49\textwidth]{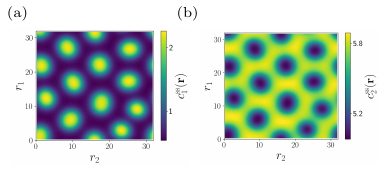}
    \caption{
    Steady state concentration fields of 
    (a)~$\ch{X_1}$ and
    (b)~$\ch{X_2}$
    of the CRN~\eqref{eq:crn-brusselator}. 
    Simulation parameters in arbitrary units: 
     $RT = 1$,
    $\chi=0$, $k_1=0.1$, $k_2=0.05$, 
    $\mu_{x}^{\circ}=0$, $\forall x$, $\mu_{Y_1}=\ln 2$, $\mu_{Y_2}=\ln 5$, $\mu_{Y_3}=\ln 0.01$, $\mu_{Y_4}=\ln 0.1$, 
    $A_\rct = 1$ $\forall \rct$, 
    $D_1=1$, $D_2=10$, $D_\text{nr}=1$, 
    $c_\text{1}^{\mathrm{h}}= 1.05$, $c_\text{2}^{\mathrm{h}}= 5.76$, $c_\text{nr}^{\mathrm{h}}=1.00$. 
    }
    \label{fig:Brusselator-conc}
\end{figure}

\begin{figure*}
    \centering
    \includegraphics[width=0.98\textwidth]{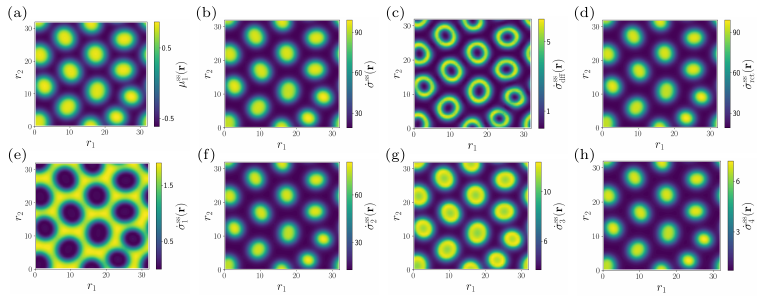}
    \caption{
    Steady-state thermodynamic properties of the CRN~\eqref{eq:crn-brusselator}. 
    (a)~Chemical potential of $\ch{X_1}$
    (b)~Density of entropy production rate $\deprss = \deprdffss + \deprrctss$.
    (c)~Density of diffusion entropy production rate $\deprdffss$.
    (d)~Density of reaction entropy production rate $\deprrctss = \sum_{\rct = 1}^4 \deprrctrss$ 
    and (e)-(f)-(g)-(h) its decomposition into the contribution of each reaction  $\rct = 1,\,2,\,3,\,4$ as explained in Subs.~\ref{subs:thermo_cost}.
    Simulation parameters are the same as in Fig.~\ref{fig:Brusselator-conc}.}
    \label{fig:Brusselator-thermo}
\end{figure*}

We solve the RD dynamics~\eqref{eq:rdeq}
when the mean-field molecular interactions and the cost of forming interfaces are given by the 
same matrixes as in Eq.~\eqref{eq:interaction-matrices-pdb}.
The typical steady-state solution and its thermodynamic properties 
are shown in Figs.~\ref{fig:Brusselator-conc} and~\ref{fig:Brusselator-thermo}, respectively.
We observe that the concentration fields spatially organize (Fig.~\ref{fig:Brusselator-conc}) into a pattern 
that is striking different from the previous ones examined in
Subs.~\ref{subs:ex_pdb_m} and~\ref{subs:example_cb_solution}
(see Figs.~\ref{fig:pdb} and~\ref{fig:cb-solution}, respectively).
In this case, \ch{X_1} accumulates in \textit{multiple droplets}
and \ch{X_2} remains almost homogeneous.
Furthermore, both reactions and diffusion processes are out of equilibrium 
since the chemical potentials are not homogeneous (Fig.~\ref{fig:Brusselator-thermo}a):
$\deprrctss>0$ (Fig.~\ref{fig:Brusselator-thermo}d) and $\deprdffss>0$ (Fig.~\ref{fig:Brusselator-thermo}c).
This has two main consequences.
First,
the dissipation due to the chemical reactions, quantified by $\deprrct$, is not homogeneous either:
reactions dissipate more inside the droplets (Fig.~\ref{fig:Brusselator-thermo}d).
However, not all reactions contribute in the same way to the total reaction entropy production rate.
Indeed, as show in Fig.~\ref{fig:Brusselator-thermo}e-h, 
reaction $\rct = 1$ dissipates more outside the droplets, 
while $\rct = 2$, $3$, and $4$ dissipate more inside the droplets. 
Furthermore, $\rct = 2$ (representing an autocatalytic reaction) is the reaction dissipating the most.
{This physically implies that reactions $\rct = 2$, $3$, and $4$  are more out of equilibrium inside the droplets
or, equivalently, they occur in a preferential direction (either forward or backward) more inside the droplets than outside
(since the entropy production rate is related to the asymmetry in the reaction fluxes via Eq.~\eqref{eq:eprrct2}).}
Second,
the diffusion processes are out of equilibrium 
and dissipate mostly at the interface between the droplets and the bulk
as shown in Fig.~\ref{fig:Brusselator-thermo}c.
{This physically means that diffusion currents are maintained with a preferential direction
at the interface between the droplets and the bulk
in such a way as to balance the consumption/production of species inside the droplets due to the chemical reactions
and sustain the steady state pattern.
}

Note that the steady-state density of nonconservative work 
(which satisfies $\dncwrkss = \deprss$) is peaked inside the droplets (see Fig.~\ref{fig:Brusselator-thermo}b) 
and, therefore, most of the free energy maintaining nonequilibrium self-organization
is provided from inside the droplets (even if the chemostatted species are homogeneous).



\section{Conclusions\label{sec:conc}}

We provided a general thermodynamic framework to study the energetics of self-organization in open non-ideal RD systems. 
We showed that it can be used to study in great detail, 
with spatial resolution, 
the contribution of the reactive and diffusive processes to the dissipation of self-organized structures.  
For special classes of RD systems, our framework allowed us to construct potentials that are minimized by the dynamics (Lyapunov functions) and whose minima correspond to the steady state.
Thermodynamic potentials (independent of the kinetics) are found for detailed balance RD systems that relax to equilibrium, while kinetic potentials are found for pseudo detailed balanced and complex balanced CRNs which relax to nonequilibrium steady states.
In doing so, we demonstrated that thermodynamics is useful not only to study the energetics of RD systems but also to make inferences on their dynamics. 

{Our framework addresses the questions posed in the introduction 
concerning the mechanisms sustaining and controlling self-organization. 
It reveals how the interplay between passive and active mechanisms 
is intricately tied to the topology of the underlying CRN.
In particular, the mechanisms are purely passive in detailed balanced CNRs (Sec.~\ref{sub:slaw}), 
where both chemical reactions and diffusion equilibrate.
In contrast, 
pseudo detailed balanced (Subs.~\ref{sub:pdb}) and complex balanced CRNs (Subs.~\ref{sub:cb}) 
exhibit a mix of active and passive traits:
chemical reactions dissipate (i.e., they are active), but diffusion equilibrates (i.e., it is passive).
In CRNs that are neither pseudo detailed balanced nor complex balanced, both chemistry and diffusion dissipate, and self-organization is purely active.
Our framework further uncovers how the control of self-organized structures 
by chemical reactions is also strongly dependent on the network topology.
In the case of pseudo detailed balanced and complex balanced CRNs 
the chemical reactions only determine the total concentrations of the species,
which diffusion then spatially organizes (Subs.~\ref{subs:diff_induced_emergence_self_org}).
Instead, chemical reactions can directly affect the spatial organization of CRNs 
that are neither pseudo detailed balanced nor complex balanced.
}

{Our framework is general and applicable to a wide range of systems in both biology and synthetic chemistry.
It can distinguish between active and passive mechanisms solely on the basis of the topology of the CRN.
However, determining the dissipation of self-organized structures requires knowledge of all diffusion and reaction fluxes, 
which can be challenging to obtain.
From this perspective,  our thermodynamic framework demands a more detailed system characterization compared to dynamical frameworks that primarily focus on reproducing self-organized structures.
}

{Furthermore, we emphasize that the crucial feature of our framework is thermodynamic consistency.
This is ensured by the local detailed balance condition,
which corresponds to the macroscopic limit~\cite{falasco2024macroscopic} 
of the local detailed balance condition of stochastic thermodynamics~\cite{Pigolotti:ST} 
when specialized for CNRs~\cite{Rao2018b,postmodernthermo2023}.
However, it is important to note that this condition might not hold at scales such as those of 
predator-prey distributions, 
vegetation patterns,
and spiral patterns of galaxies.
Hence, the direct application of our framework to these systems is not straightforward  
and the development of thermodynamically-consistent coarse-graining techniques,
like those in Refs.~\cite{Avanzini2020b,Avanzini2023},
will be necessary.

Finally, we stress that our local detailed balance condition~\eqref{eq:ldb} 
}
is in agreement with the general derivation in Ref.~\cite{Groot1984}, 
where the thermodynamic forces conjugated to the reaction currents are 
the (stoichiometric-weighted combinations of) chemical potentials $\{\cp\}$.
Other works have used instead the exchange chemical potentials $\{ \cpex\}$, 
namely, the differences between the chemical potentials and the chemical potential of another species chosen as reference.
This is only consistent with the local detailed balance condition when dealing with unimolecular (or pseudo-unimolecular) reactions.

Future studies will be needed to extend our framework to fluctuating RD systems~\cite{Tiani2023} and to explore the phenomenology of self-organized structures in non-ideal RD systems following the line of Refs.~\cite{menou2023physical,luo2023influence}. Another interesting extension concerns microswimmers which undergo a self-propelled motion~\cite{Gaspard2019, Markovich2021, Cates2022, Speck2024, Golestanian2024}; here, the challenge is to understand how to properly describe the coupling between chemical reactions, occurring either at the surface of (e.g., Janus colloids in a fuel bath) or within (e.g., bacteria fulled by nutrient consumption) microswimmers, and the resulting self-propelled dynamics.


\section{Acknowledgments}
FA is supported by the project P-DiSC\#BIRD2023-UNIPD 
funded by the Department of Chemical Sciences of the University of Padova (Italy).
TA, \'EF and ME are supported by the Fond National de la Recherche—FNR, Luxembourg: 
T.A. by the Project ThermoElectroChem (C23/MS/18060819), 
\'EF by the Project SMAC (14389168),
and ME by the Project ChemComplex (C21/MS/16356329).

\section*{Data Availability}
The data that support the findings of this study are available from the corresponding author upon reasonable request.


\appendix

\section{Minimization of the Free Energy\label{app:free_var}}
We prove here that the equilibrium steady state $\conveq$ of closed RD systems (defined in Eq.~\eqref{eq:teq})
corresponds to the minimum $\convm$ of the free energy~\eqref{eq:free} 
satisfying the conserved quantities of the RD dynamics~\eqref{eq:rdeq}
(i.e., the abundances of the moieties~\eqref{eq:cqua} as discussed in Subs.~\ref{sub:claw}).
 
We start by recognizing that 
$\convm$ is defined by the minimum $(\convm, \lmvm)$ of the Lagrangian functional
\begin{equation}\small
\begin{split}
\lagc{}= 
\free
&- \sum_{\ilaw}\lmcl\Big(\sum_{\spe}\claw \int_V\dd\pos\, \conc - \cquaref\Big)\,,
\end{split}
\end{equation}
with $\lmv = (\dots,\lmcl,\dots)$ being the vector of the Lagrange multipliers
and $\{\cquaref\}$ being the actual values of the moiety abundances.
Namely, $(\convm, \lmvm)$ is the solution of the following equations  
\begin{subequations}\small
\begin{align}
\Fd{\lagc}\bigg|_{\mini}&= \cpm -  \sum_{\ilaw} \lmclm \claw= 0\,,\label{eq:eqvarclosed1}\\
\Fdlmcl{\lagc}\bigg|_{\mini}&=-\Big(\sum_{\spe}\claw \int_V\dd\pos\, \concm - \cquaref \Big) = 0\,.
\end{align}\label{eq:eqvarclosed}%
\end{subequations}
Equation~\eqref{eq:eqvarclosed} implies that $\convm$ satisfies the equilibrium conditions in Eq.~\eqref{eq:teq} 
and, consequently, $\convm = \conveq$.
Indeed, using Eq.~\eqref{eq:eqvarclosed1}, 
\begin{equation}
\sum_{\spe} \cpm \matSi{} =  \sum_{\spe}\big(\sum_{\ilaw} \lmclm \claw\big)  \matSi{}  = 0\,,
\end{equation}
by definition of conservation laws (given in Eq.~\eqref{eq:claw}),
and $\{\cpm\}$ are homogeneous implying that Eq.~\eqref{eq:teqdff} is satisfied too.

\section{Minimization of the Semigrand Free Energy\label{app:sgfree_var}}
We prove here that the equilibrium steady state $\conveq$ of open RD systems (defined in Eq.~\eqref{eq:teq})
corresponds to the minimum $\convm$ of the semigrand free energy~\eqref{eq:sgfree} 
satisfying the conserved quantities of the RD dynamics~\eqref{eq:rdeq}
(i.e., abundances of the moieties corresponding to the unbroken conservation laws~\eqref{eq:dtcquaub} 
as discussed in Subs.~\ref{sub:claw}).
We also prove that the equilibrium chemical potentials of the $\setspeyp$ and $\setspeyf$ species
are solely determined by the reference chemical potentials $\{\cprypref\}$.

We start by recognizing that
$\convm$ is defined by the minimum $(\convm, \lmvm)$ of the Lagrangian functional 
%
\begin{equation}\small
\begin{split}
\lagc{}= &
\overbrace{\free - \sum_{\speyp} \cprypref \,\cmoi}^{=\sgfree} \\
&- \sum_{\ilawu}\lmclu\Big(\sum_{\spe}\clawu \int_V\dd\pos\, \conc - \cquauref\Big) \,,
\end{split}\label{eq:lagrangian_open}
\end{equation}
with $\lmv = (\dots,\lmclu,\dots)$ being the vector of the Lagrange multipliers 
and $\{\cquauref\}$ being the actual values of the abundances of the moieties corresponding to the unbroken conservation laws.
Namely, $(\convm, \lmvm)$ is the solution of the following equations   
\begin{subequations}\small
\begin{align}
\Fd{\lagc}\bigg|_{\mini}=& 
\cpm 
-  \sum_{\speyp,\ilawb}\cprypref  \iclawbp   \clawb - \sum_{\ilawu} \lmclum \clawu = 0 \,,
\label{eq:eqvaropen0} \\ 
\Fdlmclu{\lagc}\bigg|_{\mini}=& - \Big(\sum_{\spe}\clawu \int_V\dd\pos\, \concm - \cquauref\Big)= 0\,,
\label{eq:eqvaropen4}
\end{align}\label{eq:eqvaropen}%
\end{subequations}
where we used Eq.~\eqref{eq:moiety}.
Equation~\eqref{eq:eqvaropen0} implies that the chemical potentials in the minimum $\convm$ of the semigrand free energy~\eqref{eq:sgfree} read
\begin{equation}
\cpm = \sum_{\speyp,\ilawb}\cprypref \, \iclawbp \,  \clawb +  \sum_{\ilawu} \lmclum \clawu\,,
\label{eq:eqvaropenfinal}
\end{equation}
and, therefore, $\convm$ satisfies the equilibrium conditions in Eq.~\eqref{eq:teq}, i.e., $\convm = \conveq$.
Indeed, 
\begin{equation}
\begin{split}
\sum_{\spe} \cpm \matSi{} =  
&\sum_{\spe} \big( \sum_{\speyp,\ilawb}\cprypref \, \iclawbp \,  \clawb \big)  \matSi{}  \,
\\
&+  \sum_{\spe} \big(  \sum_{\ilawu} \lmclum \clawu \big)  \matSi{}  = 0\,,
\end{split}
\end{equation}
by definition of conservation laws (given in Eq.~\eqref{eq:claw}),
and $\{\cpm\}$ are homogeneous implying that Eq.~\eqref{eq:teqdff} is satisfied too.

\remark
According to Eq.~\eqref{eq:eqvaropenfinal},
the equilibrium chemical potentials of the internal species are determined by
i) the reference chemical potentials $\{\cprypref\}$ and 
ii) the abundances $\{\cquau\}$ of the moieties corresponding to the unbroken conservation laws via $\{\lmclum\}$.

\remark
According to Eq.~\eqref{eq:eqvaropenfinal},
the equilibrium chemical potentials of the potential and force species are solely determined by 
the reference chemical potentials $\{\cprypref\}$, according to
\begin{subequations}
\begin{align}
\cpypeq& = \cprypref \,,\\
\cpyfeq &= \sum_{\speyp,\ilawb}\cprypref \, \iclawbp \,  \clawbf \,,
\end{align}
\end{subequations}
using $\sum_{\ilawb} \iclawbpp \,  \clawbp = \dk_{\speyp, \speypp}$ and $\law{\ilawu}{\spey} = 0$.
Thus, if the force species were chemostatted in such a way that 
$\cpryf \neq \cpyfeq = \sum_{\speyp,\ilawb}\cprypref \, \iclawbp \,  \clawbf$, 
open RD systems would not admit an equilibrium steady state and detailed balance would be broken 
(as discussed in Subs.~\ref{sub:slaw} using the formulation of the second law given in Eq.~\eqref{eq:slaw}).
Similarly, 
if the potential chemostats imposed different values of the chemical potentials $\{\cpryp\}$ in different $\pos\in V$, 
open RD systems would not admit an equilibrium steady state and detailed balance would be broken.
If this is the case, we recall that the reference chemical potential $\{\cprypref\}$ entering the definition of semigrand free energy~\eqref{eq:sgfree} 
can be chosen arbitrarily among these values.
This is equivalent to choosing a reference equilibrium steady state $\conveq$ 
to which the open RD system would relax if $\cpryp = \cprypref$ and the force species were not chemostatted.

\section{Minimization of the Free Energy in Purely Diffusive Systems \label{app:dff_var}}
We prove here that the steady state $\convss$ of 
either pseudo detailed balanced CRNs (defined in Subs.~\ref{sub:pdb}) or complex balanced CRNs (defined in Subs.~\ref{sub:cb})
corresponds to the minimum $\convmm$ of the free energy $\free$ in Eq.~\eqref{eq:free} 
when the total abundances of the chemical species are fixed and equal to $\int_V\dd\pos\, \convss$.

In what follows, we use $\freeshift$ for the kinetic potential of either pseudo detailed balanced or complex balanced CRNs.
Namely, the potential obtained from the free energy~\eqref{eq:free}
after applying the shifts $\{\shx\}$ of the standard chemical potentials of the internal species 
in Eq.~\eqref{eq:shift:pdb} for pseudo detailed balanced CRNs
or in Eq.~\eqref{eq:shift:cb} for complex balanced CRNs.
Furthermore, like in Subs.~\ref{sub:pdb} and Subs.~\ref{sub:cb}, 
we assume that the chemostatted species are ideal, homogeneous and constant in time.

We have to start by recognizing that, 
in agrement with Subs.~\ref{sub:pdb} and Subs~\ref{sub:cb},
$\convss$ is a minimum of the kinetic potential $\freeshift$
satisfying the conserved quantities of the RD dynamics~\eqref{eq:rdeq}
(i.e., abundances of the moieties corresponding to the unbroken conservation laws~\eqref{eq:dtcquaub}).
Indeed, the minimum $(\convm, \lmvm)$ of the Lagrangian functional
\begin{equation}\small
\lagc{}= 
\freeshift
- \sum_{\ilawu}\lmclu\Big(\sum_{\spex}\clawux \int_V\dd\pos\, \conx - \cquauref\Big)\,,
\end{equation}
with $\lmv = (\dots,\lmclu,\dots)$ being the vector of the Lagrange multipliers 
and $\{\cquauref\}$ being the actual values of the moieties abundances corresponding to the unbroken conservation laws,
is the solution of the following equations
\begin{subequations}
\begin{align}
\Fdx{\lagc}\bigg|_{\mini}&= \cpxm + \shx - \sum_{\ilawu}\lmclum\clawux = 0\label{eq:eqvarpdbcb1}\,,\\
\Fdlmcl{\lagc}\bigg|_{\mini}&= -\Big(\sum_{\spex}\clawux \int_V\dd\pos\, \conxm - \cquauref \Big) = 0\,,
\end{align}
\label{eq:eqvarpdbcb}%
\end{subequations}
where we used the fact that the chemostatted species are ideal, homogeneous and constant in time.
Equation~\eqref{eq:eqvarpdbcb1} implies that 
$\convm$ satisfies the conditions 
in Eq.~\eqref{eq:pseudo_eq2} (resp. Eq.~\eqref{eq:complex_eq2})
for the steady state of pseudo detailed balanced (resp. complex balanced) CRNs 
with $\{\shx\}$ given in Eq.~\eqref{eq:shift:pdb} (resp. Eq.~\eqref{eq:shift:cb}).

We now consider the minimum of the free energy $\free$ in Eq.~\eqref{eq:free} 
when the total abundances of the chemical species are fixed and equal to $\int_V\dd\pos\, \convss$:
the minimum $(\convmm, \lmvmm)$ of the Lagrangian functional
\begin{equation}
\lagc{}= 
\free
- \sum_{\spex} \lmx \int_V\dd\pos\,\big( \conx - \conxss\big)\,,
\end{equation}
with $\lmv = (\dots,\lmx,\dots)$ the vector of the Lagrange multipliers, satisfying
\begin{subequations}
\begin{align}
\Fdx{\lagc}\bigg|_{\minii}&= \cpxmm -\lmxmm = 0\label{eq:vardff1}\,,\\
\Fdlmx{\lagc}\bigg|_{\minii}&= -\int_V\dd\pos\,\big( \conxmm - \conxss\big) = 0\label{eq:vardff2}\,,
\end{align}
\label{eq:vardff}%
\end{subequations}
where we used again the fact that the chemostatted species are ideal, homogeneous and constant in time.
By comparing Eq.~\eqref{eq:eqvarpdbcb} and Eq.~\eqref{eq:vardff}, 
we can conclude that Eq.~\eqref{eq:vardff} admits a solution such that 
$\convmm = \convm = \convss$ and $\lmxmm = \shx - \sum_{\ilawu}\lmclum\clawux$ $\forall\spex$.
Namely, the $\pos$-dependence of $\convss$ for
either pseudo detailed balanced CRNs or complex balanced CRNs
can be obtained by minimizing the free energy $\free$ in Eq.~\eqref{eq:free} 
with the constraint that the total abundances of the chemical species are equal to $\int_V\dd\pos\, \convss$.

\text{ }\newpage
\bibliography{biblio}
\end{document}